\documentclass[sigconf]{acmart}
\usepackage{booktabs}
\usepackage{multirow}
\usepackage{graphicx}

\usepackage{natbib}
\usepackage[all]{nowidow}
\usepackage{url}

\usepackage{xcolor}

\usepackage{layouts}
\usepackage{nth}

\AtBeginDocument{%
  \providecommand\BibTeX{{%
    \normalfont B\kern-0.5em{\scshape i\kern-0.25em b}\kern-0.8em\TeX}}}

\usepackage{amsmath}
\usepackage{caption}
\usepackage{subcaption}
\usepackage{graphicx}

\DeclareMathOperator*{\argmin}{arg\,min}
\DeclareMathOperator*{\argmax}{arg\,max}
\DeclareMathOperator{\EX}{\mathbb{E}}
\newcommand{\ind}{\mathbbm{1}}
\usepackage{mathtools}
\usepackage{bbm}
\usepackage{mismath}

\copyrightyear{2021}
\acmYear{2021}
\setcopyright{rightsretained}
\acmConference[AIES '21]{Proceedings of the 2021 AAAI/ACM Conference on AI, Ethics, and Society}{May 19--21, 2021}{Virtual Event, USA}
\acmBooktitle{Proceedings of the 2021 AAAI/ACM Conference on AI, Ethics, and Society (AIES '21), May 19--21, 2021, Virtual Event, USA}
\acmDOI{10.1145/3461702.3462571}
\acmISBN{978-1-4503-8473-5/21/05}

\author{Umang Bhatt$^{1,2}$, Javier Antor\'an$^{2}$, Yunfeng Zhang$^{3}$, Q. Vera Liao$^{3}$, Prasanna Sattigeri$^{3}$, Riccardo Fogliato$^{1,4}$, Gabrielle Gauthier Melan\c{c}on$^{5}$, Ranganath Krishnan$^{6}$, Jason Stanley$^{5}$, Omesh Tickoo$^{6}$, Lama Nachman$^{6}$,
Rumi Chunara$^{7}$, Madhulika Srikumar$^{1}$,
Adrian Weller$^{2,8}$,
Alice Xiang$^{1,9}$
}
\affiliation{
  \institution{$^{1}$Partnership on AI, $^{2}$University of Cambridge, $^{3}$IBM Research, $^{4}$Carnegie Mellon University,  $^{5}$Element AI, $^{6}$Intel Labs, $^{7}$New York University, $^{8}$The Alan Turing Institute, $^{9}$Sony AI}
}



\begin{document}
\title[Uncertainty as a Form of Transparency]{Uncertainty as a Form of Transparency: Measuring, Communicating, and Using Uncertainty}

\renewcommand{\shortauthors}{Bhatt, Antor\'an, Zhang, et al.}

\begin{abstract}
Algorithmic transparency entails exposing system properties to various stakeholders for purposes that include understanding, improving, and contesting predictions. Until now, most research into algorithmic transparency has predominantly focused on explainability. Explainability attempts to provide reasons for a machine learning model's behavior to stakeholders. However, understanding a model's specific behavior alone might not be enough for stakeholders to gauge whether the model is wrong or lacks sufficient knowledge to solve the task at hand. In this paper, we argue for considering a complementary form of transparency by estimating and communicating the uncertainty associated with model predictions. First, we discuss methods for assessing uncertainty. Then, we characterize how uncertainty can be used to mitigate model unfairness, augment decision-making, and build trustworthy systems. Finally, we outline methods for displaying uncertainty to stakeholders and recommend how to collect information required for incorporating uncertainty into existing ML pipelines. This work constitutes an interdisciplinary review drawn from literature spanning machine learning, visualization/HCI, design, decision-making, and fairness. We aim to encourage researchers and practitioners to measure, communicate, and use uncertainty as a form of transparency.
\end{abstract}

\maketitle

\section{Introduction}
Transparency in machine learning (ML) encompasses a wide variety of efforts to provide stakeholders, such as  model developers and end users, with relevant information about how a ML model works~\citep{o2018linking,weller2019transparency,bhatt2020explainable}.
One form of transparency is procedural transparency, which provides information about model development (e.g., code release, model cards, dataset details)~\citep{gebru2018datasheets,raji2019ml,arnold2019factsheets,mitchell2019model}. Another form is algorithmic transparency, which exposes information about a model's behavior to various stakeholders~\cite{ribeiro2016should,sundararajan2017axiomatic,koh2017understanding}. The ML community has mostly considered explainability, which attempts to provide reasoning for a  model's behavior to stakeholders, as a proxy for algorithmic transparency~\citep{lucic2021multistakeholder}. With this work, we seek to encourage researchers to study uncertainty as an alternative form of algorithmic transparency and practitioners to communicate uncertainty estimates to stakeholders.
Uncertainty is crucial yet often overlooked in the context of ML-assisted, or automated, decision-making~\citep{schum2014toward,kochenderfer2015decision}.
If well-calibrated and effectively communicated, uncertainty can help stakeholders understand when they should trust model predictions and help developers address fairness issues in their models~\citep{tomsett2020rapid, zhang2020effect}.

Uncertainty refers to our lack of knowledge about some outcome. As such, uncertainty will be characterized differently depending on the task at hand. 
In regression tasks, uncertainty is often expressed in terms of error bars, also known as confidence intervals.
For example, when predicting the number of crimes in a given city, we could report that 
the number of predicted crimes is $943 \pm 10$, where
``$\pm 10$'' represents a 95\% 
\textit{confidence interval} (capturing two standard deviations on either side of the central, mean estimate). 
The smaller the interval, the more certain the model. 
In classification tasks, probabilities are often used to capture how confident a model is in a specific prediction.
For example, a classification model may decide that a person is at a high risk for developing diabetes given a prediction of $85$\% chance of diabetes. 
Broadly, uncertainty in data-driven decision-making systems may stem from different sources and thus communicate different information to stakeholders~\citep{hora1996aleatory,gal2016uncertainty}. 
Aleatoric uncertainty is induced by inherent randomness (or noise) in the quantity to predict given inputs. 
Epistemic uncertainty can arise due to lack of sufficient data to learn our model precisely. 

\subsection*{Why do we care?}
We posit uncertainty can be useful for obtaining
fairer models, improving decision-making, and building trust in automated systems.
Throughout this work, we will use the following cancer diagnostics scenario for illustrative purposes:
Suppose we are tasked with diagnosing individuals as having breast cancer or not, as in~\cite{curtis2012genomic,dua}. Given categorical and continuous characteristics about an individual (medical test results, family medical history, etc.), we estimate the probability of an individual having breast cancer. We can then apply a threshold to classify them into high- or low-risk groups. Specifically, we have been tasked with building ML-powered tools to help three distinct audiences: doctors, who will be assisted in making diagnoses; patients, who will be helped to understand their diagnoses; and review board members, who will be aided in reviewing doctors' decisions across many hospitals. 
Throughout the paper, we will refer back to the scenario above to discuss how uncertainty may arise in the design of an ML model and why, when well-calibrated and well-communicated, uncertainty can act as a useful form of transparency for stakeholders. We now briefly examine how ignoring uncertainty can be detrimental to transparency in three different use cases with the help of our scenario.

\textbf{Fairness}:
Uncertainty, if not properly quantified and considered in model development, can endanger efforts to assess model fairness. Developers often aim to assess fairness to mitigate or prevent unwanted biases.  
Breast cancer is more common for older patients, potentially leading to datasets where young age groups are underrepresented.
If our breast cancer diagnostic tool is trained on data presenting such a bias, the model might be under-specified and present larger error rates for younger patients. 
Such dataset bias will manifest itself as epistemic uncertainty.
In Section~\ref{sec:use}, we detail ways uncertainty interacts with bias in the data collection/modeling stages and how such biases can be mitigated by ML practitioners.

\textbf{Decision-making}:
Treating all model predictions the same, independent of their uncertainty, can lead decision-makers to over-rely on their models in cases where they produces spurious outputs or to under-rely on them when their predictions are accurate. 
As such, a doctor could make better use of an automated decision-making system by observing its uncertainty estimates before leveraging the model's output in making a diagnosis. 
In Section \ref{sec:use}, we draw upon the literature of judgment and decision-making (JDM) to discuss the potential implications of showing uncertainty estimates to end users of ML models.

\textbf{Trust in Automation}:
Well-calibrated and well-communicated uncertainty can be seen as a sign of model trustworthiness, in turn improving model adoption and user experience.
However, when communicated inadequately, uncertainty estimates can be incomprehensible to stakeholders and perceived negatively, thus spawning confusion and impairing trust formation. 
Suppose our model's predictions of a patients' breast cancer stage are accompanied by 95\% confidence intervals. Due to the pervasiveness of breast cancer, a large number of healthy patients might fall within these errorbars, suggesting to doctors that they could have early stage breast cancer.
In this situation, the doctors may choose to always override the model's output, the model's seeming imprecision resulting in an erosion of the doctors' trust. 
For this task, errors bars representing the predictive interquartile range might have been more appropriate. 
The optimal way to communicate uncertainty will depend on the task at hand. 
In Section \ref{sec:use}, we review prior work on how users form trust in automated systems and discuss the potential impact of uncertainty on this trust.

This work is structured as follows.
In Section~\ref{sec:measuring}, we review possible sources of uncertainty and methods for uncertainty quantification.
In Section~\ref{sec:use}, we describe how uncertainty can be leveraged in each use case mentioned above.
Finally, in Section~\ref{sec:communication}, we discuss how to communicate uncertainty effectively. Therein, we also discuss how to take a user-centered approach to collecting requirements for uncertainty quantification and communication.

\section{Measuring Uncertainty}\label{sec:measuring}

In ML, we use the term uncertainty to refer to our lack of knowledge about some outcome of interest.  We use the tools of probability to reason about and quantify uncertainty.
The Bayesian school of thought interprets probabilities as subjective degrees of belief in an outcome of interest occurring \citep{mackay2003information}. 
For frequentists, probabilities reflect how often we would observe the outcome if we were to repeat our observation multiple times \citep{Bland1151,bayes_vs_frequentist}. Fortunately for end-users, uncertainty from Bayesian and frequentist methods conveys similar information in practice \citep{frequentist_paramerter_estimation}, and can almost always be treated interchangeably in downstream tasks.

\begin{figure}[]
     \centering
     \includegraphics[width=\columnwidth]{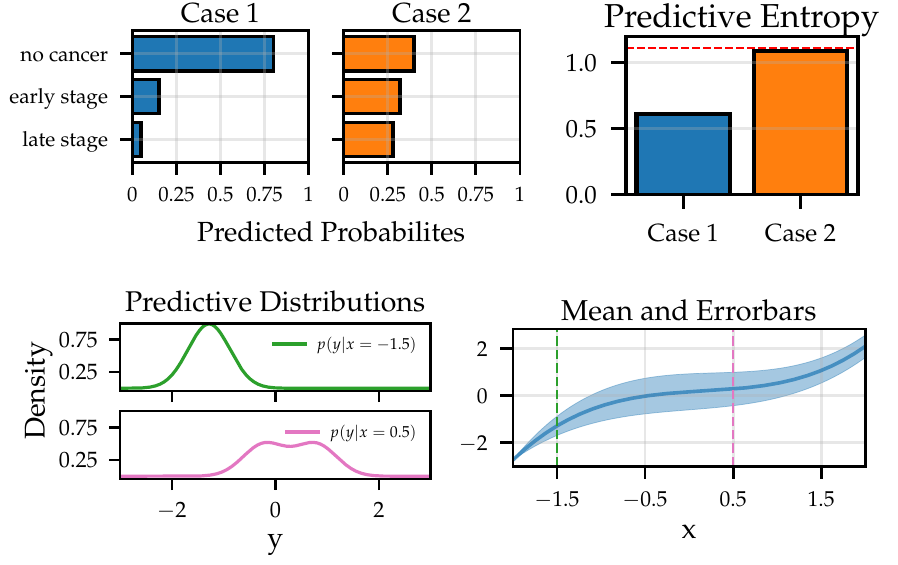}
    \caption{Top: The same prediction (no cancer) is made in two different hypothetical cancer diagnosis scenarios. However, our model is much more confident in the first. This is reflected in the predictive entropy for each case (dashed red line denotes the maximum entropy for 3-way classification). Bottom: In a regression task, a predictive distribution contains rich information about our model's predictions (modes, tails, etc). We summarise predictive distributions with means and error-bars (here standard deviations). \label{fig:uncertainty_metrics_plot}}
\end{figure}  

\subsection{Metrics for Uncertainty}
The metrics used to communicate uncertainty vary between research communities and application domains.
Predictive distributions, shown in Figure~\ref{fig:uncertainty_metrics_plot}, tell us about our models' degree of belief in every possible prediction. Despite containing a lot of information (prediction modes, tails, etc.), a full predictive distribution may not always be desirable. In our cancer diagnostic scenario, we may want our automated system to abstain from making a prediction and instead request the assistance of a medical professional when its uncertainty is above a certain threshold.  When deferring to a human expert, it may not matter to us whether the system believes the patient has cancer or not, just how uncertain it is. For this reason, \textbf{summary statistics} of the predictive distribution are often used to convey information about uncertainty. 
For classification, the predictive distribution is composed of class probabilities. These intuitively communicate our degree of belief in an outcome. On the other hand, predictive entropy decouples our predictions from their uncertainty, only telling us about the latter.
For regression, the predictive distribution is often summarized by a predictive mean together with error bars (written $\pm \sigma$). These commonly reflect the standard deviation or some percentiles of the predictive distribution. 
In Section~\ref{sec:requirements}, we discuss how the best choice of uncertainty metric is use case dependent.
We show common summary statistics for uncertainty in Table~\ref{tab:uncertainty_metrics} and elaborate in the supplementary material.

\begin{table}[]
\caption{Commonly used metrics for the quantification and communication of uncertainty.}
\label{tab:uncertainty_metrics}
\resizebox{\columnwidth}{!}{%
\begin{tabular}{@{}ccc@{}}
\toprule
                                                     & Full Information                         & Summary Statistics                    \\ \midrule
\multicolumn{1}{c|}{\multirow{2}{*}{Regression}}     & \multirow{2}{*}{Predictive Density} & Predictive Variance, Percentile       \\
\multicolumn{1}{c|}{}                                &                                          & (or quantile) Confidence Intervals    \\
\multicolumn{1}{c|}{\multirow{2}{*}{Classification}} & \multirow{2}{*}{Predictive Probabilities}  & Predictive Entropy, Expected Entropy, \\
\multicolumn{1}{c|}{}                                &                                          & Mutual Information, Variation Ratio   \\ \bottomrule
\end{tabular}%
}
\end{table}

\subsection{The Different Sources of Uncertainty}
While there can be many sources of uncertainty \citep{van_der_bles_communicating_nodate}, herein we focus on  those that we can quantify in ML models: \textit{aleatoric uncertainty} (also known as indirect uncertainty) and \textit{epistemic uncertainty} (also known as direct uncertainty) ~\cite{der2009aleatory,gal2016uncertainty,Depeweg_thesis}.

\textbf{Aleatoric uncertainty} stems from noise, or class overlap, in our data.  Noise in the data is a consequence of unaccounted-for factors that introduce variability in the inputs or targets. Examples of this could be background noise in a signal detection scenario or the imperfect reliability of a medical test in our cancer diagnosis scenario. 
Aleatoric uncertainty is also known as irreducible uncertainty: it cannot be decreased by observing more data.
If we wish to reduce aleatoric uncertainty, we may need to leverage different sources of data, e.g., switching to a more reliable clinical test. 
In practice, most ML models account for aleatoric uncertainty through the specification of a noise model or likelihood function. 
A homoscedastic noise model makes the assumption that all of the input space is equally noisy, $y = f(x) + \epsilon; \!\!\!\!\quad \epsilon\,{\sim}\,p(\epsilon)$.  However, this may not always be true. Returning to our medical scenario, consider using results from a clinical test which produces few false negatives but many false positives as an input to our model. A heteroscedastic noise assumption allows us to express aleatoric uncertainty as a function of our inputs $y = f(x) + \epsilon;\!\!\!\!\quad \epsilon\,{\sim}\,p(\epsilon| x)$. Perhaps the most commonly used heteroscedastic noise models in ML are those induced by the sigmoid or softmax output layers. These enable almost any classification model to express aleatoric uncertainty: see the supplementary material
for more details.

\textbf{Epistemic uncertainty} stems from a lack of knowledge about which function best explains the data we have observed. There are two reasons why epistemic uncertainty may arise. Consider a scenario in which we employ a very complex model relative to the amount of training data available. We will be unable to properly constrain our model's parameters. This means that, out of all the possible functions that our model can represent, we are unsure of which ones to choose.  In this work, we refer to uncertainty about a model's parameters as \textit{model uncertainty}. We might also be uncertain of whether we picked the correct model class in the first place. Perhaps we are using a linear predictor but the phenomenon we are trying to predict is non-linear.  We will refer to this as \textit{model specification uncertainty} or \textit{architecture uncertainty}. 
Epistemic uncertainty can be reduced by collecting more data in input regions where the training dataset was sparse. It is less common for ML models to capture epistemic uncertainty. Often, those that do are referred to as probabilistic models.

\begin{figure*}
    \begin{subfigure}{0.7\textwidth}
         \centering
         \includegraphics[width=\textwidth]{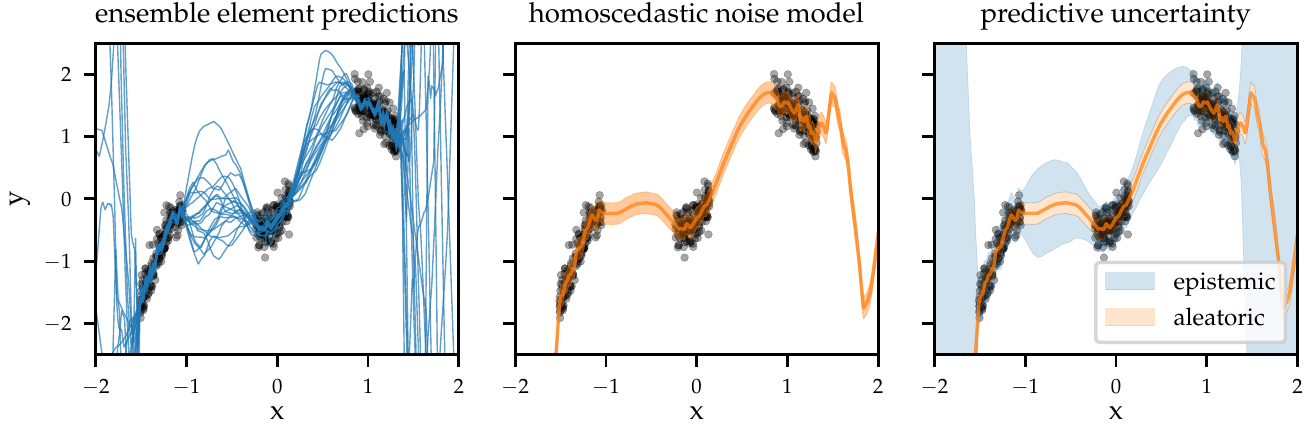}
     \end{subfigure}
     \hfill
     \begin{subfigure}{0.26\textwidth}
         \centering
         \includegraphics[width=\textwidth]{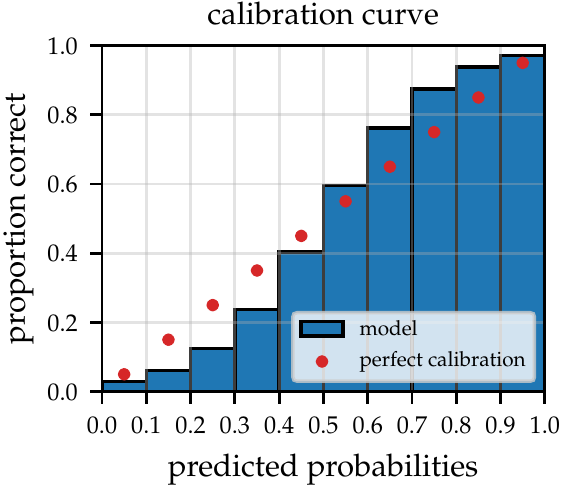}
     \end{subfigure}
        \caption{Uncertainty quantification and evaluation. Left three plots: A 15-element deep ensemble provides increased model (epistemic uncertainty) in data sparse regions. A homoscedastic Gaussian noise model provides aleatoric uncertainty matching the noise in the data. Both combine to produce a predictive distribution. Right: calibration plot for classification task. Each bar corresponds to a bin in which predictions are grouped. Their height corresponds to their proportion of correct predictions.}
        \label{fig:ensemble_calibration_plot}
\end{figure*}

Given a probabilistic predictive model, aleatoric and epistemic uncertainties can be quantified separately, as described in the supplementary material. 
We depict them separately in Figure \ref{fig:ensemble_calibration_plot}.
Being aware of which regions of the input space present large aleatoric uncertainty can help ML practitioners identify issues in their data collection process. On the other hand, epistemic uncertainty tells us about which regions of input space we have yet to learn about. Thus, epistemic uncertainty is used to detect dataset shift \citep{ovadia2019can}, or adversarial inputs \citep{Bayesian_active_learning}. It is also used to guide methods that require exploration like active learning \citep{houlsby2011bayesian}, continual learning \citep{2018variational}, Bayesian optimisation \citep{hernandez2014predictive}, 
and reinforcement learning \citep{Successor_Uncertainties}.

\subsection{Methods to Quantify Uncertainty}
Most ML approaches involve a noise model, thus capturing aleatoric uncertainty. However, few are able to express epistemic uncertainty. When we say that a method is able to quantify uncertainty, we are implicitly referring to those that capture both epistemic and aleatoric uncertainty. These methods can be broadly classified into two categories: Bayesian approaches \cite{neal_thesis,gp_book,Ghahramani2015}
and Frequentist, approaches \cite{Breiman1996,understanding_ML,lakshminarayanan2017simple}

\textbf{Bayesian methods} explicitly define a hypothesis space of plausible models \textit{a priori} (before observing any data) and use deductive logic to update these priors given the observed data. 
In parametric models, like Bayesian Neural Networks (BNNs)~\cite{mackay1992practical,neal1995bayesian}, this is most often done by treating model weights as random variables instead of single values, and assigning them a prior distribution $p(\mathrm{w})$.
Given some observed data $\mathrm{D} = \{\mathrm{y}, \mathrm{x}\}$, the conditional likelihood $p(\mathrm{y}\,|\,\mathrm{x}, \mathrm{w})$ tells us how well each weight setting $\mathrm{w}$ explains our observations. The likelihood is used to update the prior, yielding the posterior distribution over the weights $p(\mathrm{w}|\mathrm{D})$:
\begin{equation}
p(\mathrm{w}|\mathrm{D}) = \frac{p(\mathrm{y}\,|\,\mathrm{x},\mathrm{w})\,p(\mathrm{w})}{\int{p(\mathrm{y}\,|\,\mathrm{x},\mathrm{w})\,p(\mathrm{w})}\,\mathrm{dw}}
\label{eq:bayes}
\end{equation}
Prediction for a test point $\mathrm{x}^{*}$ is made via marginalization: all possible weight configurations are considered with each configuration's prediction being weighed by that weights' posterior density. The disagreement among predictions from different plausible weight settings induces model (epistemic) uncertainty. The predictive posterior distribution:
\begin{gather} \label{eq:predicitve_posterior}
    p(\mathrm{y} | \mathrm{x}^{*}) = \int p(\mathrm{y} | \mathrm{x}^{*}, \mathrm{w}) p(\mathrm{w}|\mathrm{D})\,d\mathrm{w}
\end{gather}
captures both epistemic and aleatoric uncertainty. 

Recently, the ML community has moved towards favoring NNs as their choice of model due to their flexibility and scalability to large amounts of data.
Unfortunately, the more complicated the model, the more difficult it is to compute the exact posterior $p(\mathrm{w}|\mathrm{D})$ and predictive $p(\mathrm{y} | \mathrm{x}^{*})$ distributions. 
For NNs, it is analytically and computationally intractable~\citep{hernandez2015probabilistic}. However, various approximations have been proposed. Among the most popular are variational inference~\cite{hinton1993keeping,blundell2015weight,gal2016dropout} and stochastic gradient MCMC~\cite{welling2011bayesian,chen2014stochastic,zhang2020csgmcmc}. 
Methods that provide more faithful approximations, and thus more calibrated uncertainty estimates, tend to be more computationally intensive and scale worse to larger models. As a result, the best method will vary depending on the use case.
Also worth mentioning are Bayesian non-parametrics, such as Gaussian Processes \citep{gp_book}. However, for brevity, we discuss these, with a broader range of Bayesian methods, in the supplementary material.

    \textbf{Frequentist methods} do not specify a prior distribution over hypothesis. They exclusively consider how well the distribution over observations implied by each hypothesis matches the data. Here, uncertainty stems from how we expect our chosen hypothesis to change if we were to repeatedly sample different sets of data.
    Perhaps the most salient Frequentist technique is ensembling~\cite{ensemble_methods,Lobato2009PredictionBO}. This consists of training multiple models in different ways to obtain multiple plausible fits. At test time, the disagreement between ensemble elements' predictions yields model uncertainty, as show in Figure \ref{fig:ensemble_calibration_plot}. 
    Currently, deep ensembles \cite{lakshminarayanan2017simple} are one of best performing uncertainty quantification approaches for NNs \citep{ashukha2019pitfalls}, retaining calibration even under dataset shift~\cite{ovadia2019can}. Unfortunately, the computational cost involved with running multiple models at both train and test time also make ensembles one of the most expensive methods. There is a vast heterogeneous literature on frequentist uncertainty quantification, some of which is covered in the supplementary material. 
    
\subsection{Uncertainty Evaluation}\label{subsec:eval}
Calibration is a form of quality assurance for uncertainty estimates.
It is not enough to provide larger error bars when our model is more likely to make a mistake. Our predictive distribution must reflect the true distribution of our targets. 
Recall our cancer diagnosis scenario, where the system declines to make a prediction when uncertainty is above a threshold, and a doctor is queried instead. 
Due to the doctor's time being limited, we might design our system such that it only declines to make a prediction if it estimates there is a probability greater than $0.05$ of the prediction being wrong. 
If instead of being well-calibrated, our system is underconfident, we would over-query the doctor in situations where the AI's prediction is correct. Overconfidence would result in taking action on unreliable predictions: delivering unnecessary treatment or abstaining from providing necessary treatment.

Calibration is orthogonal to accuracy. 
A model with a predictive distribution that matches the marginal distribution of the targets $p(y | x) = p(y)\, \forall\,x$ would be perfectly calibrated but would not provide any useful predictions.  Thus, calibration is usually measured in tandem with accuracy through either a general fidelity metric (most often chosen to be a proper scoring rule \cite{gneiting2007strictly}) which subsumes both objectives, or using two separate metrics.
The most common metrics of the former category are \textit{negative log-likelihood} (NLL) and \textit{Brier score}~\cite{brier1950verification}. Of the latter type,  \textit{Expected calibration error} (ECE)~\cite{naeini2015obtaining}, illustrated in Figure \ref{fig:ensemble_calibration_plot}, is popularly used in classification scenarios. We refer to the supplementary material
for a detailed discussion of calibration metrics.

Recently, \citet{antoran2020getting} introduced CLUE, a method to help identify which input features are responsible for models' uncertainty. This class of approach opens avenues for stakeholders to qualitatively evaluate if their models' reasons for uncertainty align with their own intuitions~\citep{ley2021delta}. 
A transparent ML model requires both well-calibrated uncertainty estimates and an effective way to communicate them to stakeholders.  
If uncertainty is not well-calibrated, our model cannot be transparent since its uncertainty estimates provide false information. Thus calibration is a precursor to using uncertainty as a form of transparency. 

\bigskip
\noindent\textbf{Takeaways}: 
\begin{enumerate}
    \item Well-calibrated uncertainty is a key factor in transparent  data-driven decision-making.
	\item  Uncertainty can stem from both the difficulty of the problem we are trying to solve and the modelling choices we are using to solve it.
	\item The more complex the problem, the more costly it is to obtain calibrated uncertainty estimates.
\end{enumerate}

\section{Using Uncertainty}

\label{sec:use}
In this section, we discuss the use of uncertainty based on the three motivations presented in the Introduction: fairness, decision making and trust in automation. These are not mutually exclusive; as such, they could all be leveraged simultaneously depending on the use case. In Section~\ref{sec:requirements}, we discuss the importance of gathering requirements on how stakeholders, both internal and external, will use uncertainty estimates.

\subsection{Uncertainty and Fairness}
An unfair ML model is one that exhibits unwanted bias towards or against one or more target classes. 
Here, we discuss possible ways in which bias can appear as a consequence of unaccounted-for uncertainty and potential approaches to mitigate it. 
The supplementary material contains further discussion on the interplay of model specification uncertainty and fairness.

{\it Measurement bias}, also known as feature noise, is a case of aleatoric uncertainty (defined in Section~\ref{sec:measuring}). It arises when one or more of the features in the data only represent a proxy for the features that would have ideally been measured and can be mitigated by a properly specified nose model.
We describe the potential effects of noise on inputs and targets. 

Commencing with the former, we discuss noise in the sensitive attribute.
In contexts such as medical diagnosis, information on race and ethnicity of patients may not be collected~\cite{chen2019fairness}. Alternatively, in some contexts, survey participants may have incentives to misreport characteristics such as their religious or political affiliations. 
The experimental results of~\citet{gupta2018proxy} have shown that enforcing fairness constraints on a noisy sensitive attribute, without assumptions on the structure of the noise, is not guaranteed to lead to any improvement in the fairness properties of our model. 
Successive papers have explored which assumptions are needed in order to obtain such guarantees. 

The ``mutually contaminated learning model'' assumes the noise only depends on the true unobserved value of the sensitive attribute~\citep{scott2013classification}.
Here, measures of demographic parity and equalized odds computed on the observed data are equal to the true metrics up to a scaling factor, which is proportional to the value of the noise~\citep{lamy2019noise}; if the noise rates are known, then the true metrics can be directly estimated. 
When information on the sensitive attribute is unavailable (e.g., information on gender was not collected), but can be predicted from an auxiliary dataset, disparity measures are generally unidentifiable~\citep{chen2019fairness, kallus2019assessing}.

Second, we discuss noise in the targets (i.e., aleatoric uncertainty in the labels). This source of bias has attracted less attention in the fairness community despite being similarly pervasive. \citet{obermeyer2019dissecting} found that when medical expenses were used as a proxy for illness, their algorithm severely underpredicted the prevalence of the illness for the population of Black patients.

Similarly to noise in the sensitive attribute, \citet{jiang2020identifying, blum2019recovering} show that fairness constraints are guaranteed to improve the properties of a predictor only under appropriate assumptions on the label noise. Indeed, \citet{fogliato2020fairness} show that even a small amount of label noise can greatly impact the assessment of the fairness properties of the model. 
There is a small but growing body of work on appropriate noise model specification for bias mitigation.
For example, the problem of auditing algorithms for fairness in the positive and unlabeled learning setting (i.e., noise is one-sided) has been studied by~\citet{fogliato2020fairness}.

{\it Representation bias} stems from how we define the population under study and how we sample our observations from said population \cite{suresh2019framework}. Representation bias is epistemic in nature and thus may be reduced by collecting additional, potentially more diverse, data. Models trained in the presence of representation bias could exhibit unwanted bias towards an under-represented group. For example, the differential performance of gender classification systems across racial groups may be due to under-representation of Black individuals in the sampled population~\citep{buolamwini2018gender}. Similarly, historical over-policing of some communities has unavoidable impacts on the sampling of the data used to train models for predictive policing~\citep{lum2016predict,ensign2018runaway}.

In our cancer diagnostics scenario, the data used to train a model for a specific hospital should closely reflect the demographic statistics of that hospitals patients, i.e. ensuring proper representation in terms of age groups, cancer stages, etc. As such, ideally, the data used to train our hospitals models, would stem from its own patients, and not those of another hospital where population statistics may be different~\citep{olteanu2019social}. 
 
In order to tackle representation bias, we must first detect that such bias exists. This can be done by leveraging uncertainty as a form of transparency. ML practitioners building a probabilistic model could check for representation bias by ensuring that their validation datasets closely matches the distribution expected at test time (in deployment). Their model presenting large epistemic uncertainty on this validation set would indicate the existence of representation bias in the training data. 
The practitioner could then identify which subgroups are unequally represented between their training and validation sets~\citep{antoran2020getting}. Finally, they could leverage this knowledge to improve the data collection procedure. 
Alternatively, we could consider designing a system that automatically checks for epistemic inputs in deployment, thus raining an alert if the statistics of the population to which it is exposed change over time. 

Unfortunately, sample size may still represent an issue. It is not always simple to collect more diverse data.  Additionally, in many domains, the existing sample size may also not be large enough to assess the existence of biases \cite{ethayarajh2020your}. 
We refer to \citet{Mehrabi2019ASO} for a more detailed breakdown of the potential sources of representation bias.

\subsection {Uncertainty and Decision-making}
Depending on the context, ML systems can be used to support human decision-making to various degrees or even substitute it outright. Crucially, in all types of data-driven decision-making, uncertainty plays a key role. 
Uncertainty enables stakeholders to better understand their decision-making system and thus better combine the systems' output with their own judgement.

First, we consider fully automated decision-making without human intervention. Decision theory~\citep{bishop_pattern} allows us to combine probabilistic predictions $p(\mathrm{y}|\mathrm{x})$ about an outcome of interest $\mathrm{y}$ given an input $\mathrm{x}$ with the cost of taking each action $\mathrm{a}$ given each outcome $L(\mathrm{a} | \mathrm{y})$, leading us to make an optimal decision $\mathrm{a}^{*}$ under uncertainty:
\begin{gather}\label{eq:bayesian_decision}
 \mathrm{a}^{*} = \argmin_{\mathrm{a}} \int L(\mathrm{a} | \mathrm{y}) p(\mathrm{y}|\mathrm{x})\, dy.
\end{gather}
Recall our medical scenario; our model is tasked with providing breast cancer diagnoses $p(\text{cancer} | \mathrm{x})$. Here, we might quantify the cost of a false negative $L(\text{report $=$ healthy} | \text{cancer})$ as 100 times greater than that of a false positive $L(\text{report $=$ cancer} | \text{healthy})$. Armed with this information, the optimal diagnosis we report to the patient can simply be obtained by plugging into Equation~\ref{eq:bayesian_decision}.
More reasonably, we may decide to have a medical professional intervene in a small number of situations where our system is likely to be wrong. This is known as a ``reject option’’ \cite{reject_option}. In turn, decision theory can be used to appropriately place a rejection threshold.

In both the fully automated and ``reject option’’ systems, uncertainty increases the transparency of the decision-making system. 
When a decision is made automatically, a model’s uncertainty can be leveraged post-hoc to help elucidate why that specific decision was made. 
When the reject option is triggered, learning about the models’ uncertainty (in the form of a predictive distribution or some summary statistic) may aid the human expert in her decision. 

We now consider situations of the latter sort: a human expert is tasked with making a decision while being aided by an ML system. Here, uncertainty plays a key role, as the end user has to weight how much they should trust the model's output.
This question corresponds to prototypical tasks \footnote{Note that social scientists often use the terms ``risk'' and ``uncertainty'' differently than in this text~~\cite{knight1921risk,rakow2010risk}} studied in the Judgment and Decision-Making (JDM) literature, i.e., ``action threshold decision’’ and ``multi-option choices’’~\cite{fischhoff2014communicating}. 
The ML literature has only just begun to examine how uncertainty estimates affect user interactions with ML models and task performance~\cite{arshad2015investigating,zhang2020effect}. However, we highlight relevant conclusions from the JDM literature that pertain to using uncertainty estimates in decision-making.  
Prospect Theory suggests that uncertainty (or risk) is not considered independently but together with the expected outcome~\cite{tversky_advances_1992,kahneman2013prospect}. This relationship is non-linear and  asymmetrical.
A certain prediction of a large loss is often perceived more negatively than an uncertain prediction of a small loss.
When risk is presented positively, e.g., a 95\% chance of the model being correct, people tend to be risk-averse; when it is presented negatively they are risk-seeking. 
Additionally, as the stake of the decision-outcome increases, tolerance for uncertainty seems to decrease at a superlinear rate~\cite{van2020effects}. 
Of course, differences among individuals’ tolerances for uncertainty also play an important role~\cite{miller1987monitoring,sorrentino1988uncertainty,heath1991preference,politi_communicating_2007}.

Assessment of risk, however, also depends on how uncertainty is communicated and perceived. Both lay people and experts rely on mental shortcuts, or heuristics, to \textit{interpret} uncertainty~\cite{tversky1974judgment}. This could lead to biased appraisals of uncertainty even if model outputs are well-calibrated. We discuss this in Section~\ref{sec:communication}. \citet{stowers2017insights} find that communicating and visualizing uncertainty information to operators of unmanned vehicles helped improve human-AI team performance; however, they note that their findings may not generalize to other tasks and contexts.

To our knowledge, the empirical understanding of how decision-makers make use of aleatoric versus epistemic uncertainty is limited. Furthermore, the JDM literature has mostly focused on discrete outcomes. There is not a good understanding of how people perceive uncertainty over continuous outcomes (e.g., errorbars). We judge these to be important gaps in the human-ML interaction literature.

\subsection{Uncertainty and Trust Formation}
While trust could be implicit in a decision to rely on a model's suggestion, the communication of a model's uncertainty can also affect people's general trust in an ML system. At a high level, communicating uncertainty is a form of transparency that can help gain people's trust. 
The relationship between uncertainty estimates and trust in automation is a relatively unexplored idea. Here, we explore the relevant literature on the underlying construct of trust and and the processes of trust formation with the goal of painting a more complex picture of how stakeholders might use uncertainty estimates to form trust in an ML system.

While not limited to ML systems, the HCI and Human Factors communities have a long history of studying trust in automation~\cite{lee2004trust,hoff2015trust,korber2018theoretical}.
 These models of trust often build on \citet{mayer1995integrative}'s classic ABI (Ability, Benevolence, Integrity) model of inter-personal trust.
Taking into account some fundamental differences between interpersonal trust and trust in automation, \citet{lee2004trust} adapted the ABI model to trust in automated systems. They highlight three underlying dimensions: 1) competence of the system within the targeted domain, 2) intention of developers: the extent to which they are perceived to want to do good to the trustor, and 3) predictability/understandability: the extent to which the system consistently operates according to a set of principles that the trustor finds acceptable. 

We speculate that communicating uncertainty estimates could be relevant to all three dimensions. If a model’s uncertainty is perceived to be too large or miscalibrated, it may harm the model's perceived competence. 
If uncertainty is not communicated or intentionally mis-communicated, users or stakeholders might form a negative opinion on the intention of developers.
If a model shows uncertainty that could not be understood or expected, it will be negatively perceived in predictability.

To anticipate how uncertainty estimates, and ways to communicate them, could impact stakeholder trust, we highlight existing ``process models’’ on how people develop trust.
Rooted in information-processing and decision-making theories~\cite{petty1986elaboration,chaiken1999heuristic,kahneman2011thinking}, ``process models’’ differentiate between an analytic (or systematic) process of trust formation and an affective (or heuristic) process of trust formation~\cite{lee2004trust,sundar2008main,metzger2013credibility}. 

The former involves rational evaluation of a trustee's characteristics; systematic trust formation in an ML system could be facilitated by providing detailed probabilistic uncertainty estimates.
The latter process relies on feelings or heuristics to form a quick judgment to trust or not; when lacking either the technical ability or motivation to perform an analytic evaluation, people rely more on the affective or heuristic route~\cite{petty1986elaboration,sundar2019machine}. 
For example, for some users the mere presence of uncertainty information could signal that the ML engineers are being transparent and sincere, enhancing their trust~\cite{hovland1953communication}. For others, uncertainty could invoke negative heuristics~\cite{van_der_bles_communicating_nodate}.
Furthermore, prior work suggests that the style in which uncertainty estimates are communicated is highly relevant to how these are perceived~\cite{parasuraman2004trust}. We elaborate on communication methods for uncertainty in Section~\ref{sec:communication}.

Lastly, we highlight a non-trivial point that the goal of presenting uncertainty estimates to stakeholders should support forming \textit{appropriate} trust, rather than blindly enhancing trust. A well-measured and well-communicated uncertainty estimate should not only facilitate the \textit{calibration} of overall trust on a system, but also the \textit{resolution} of trust~\cite{cohen1998trust,lee2004trust}. The latter referring to how precisely the judgment of trust could differentiate types of model capabilities in decision-making.

Based on relevant work from HCI, JDM, and ML, we argue that leveraging uncertainty as a form of transparency can be helpful for trust formation.
However, how uncertainty estimates are processed for trust formation and what kind of affective impact uncertainty could invoke remain open questions and merit future research.

\bigskip
\noindent\textbf{Takeaways}: 
\begin{enumerate}
    \item Uncertainty can manifest as unfairness in the form of noisy features/targets (aleatoric) or in the sampling procedure (epistemic). Being aware of uncertainty can allow ML practitioners to mitigate these issues.   
    \item Being aware of uncertainty allows stakeholders to make better use of ML-assisted decision-making systems.
    \item Delivering uncertainty estimates to stakeholders can enhance transparency by ensuring trust formation.
\end{enumerate}


\section{Communicating Uncertainty}
\label{sec:communication}
Treating uncertainty as a form of transparency requires accurately communicating it to stakeholders. However, even well-calibrated uncertainty estimates could be perceived inaccurately by people because (a) they have varying levels of understanding about probability and statistics, and (b) human perception of uncertainty quantities is often biased by decision-making heuristics. In this section, we will review some of the issues that hinder people's understanding of uncertainty estimates and will discuss how various communication methods may help address these issues. We will first describe how to communicate uncertainty in the form of confidence or prediction probabilities for classification tasks, and then more broadly in the form of ranges, confidence intervals, or full distributions. 
In the supplementary material, we dive into a case study on the utility of uncertainty communication during the COVID-19 pandemic.


\subsection{Issues in Understanding Uncertainty}
Many application domains involve communicating uncertainty estimates to the general public to help them make decisions, e.g., weather forecasting, 
transit information delivery \cite{kay_when_2016}, medical diagnosis and interventions \cite{politi_communicating_2007}. One key issue in these applications is that a great deal of their audience may not have the numeracy skills required to interpret uncertainty correctly. In a survey \citet{galesic_statistical_2010} conducted in 2010 on statistical numeracy across the US and Germany, they found that many people do not understand relatively simple statements that involve statistics concepts. For example, 20\% of the German and US participants could not say ``which of the following numbers represents the biggest risk of getting a disease: 1\%, 5\%, or 10\%,'' and almost 30\% could not answer whether 1 in 10, 1 in 100, or 1 in 1000 represents the largest risk. Another study found that people's numeracy skills significantly affect how well they comprehend risks~\citep{zikmund-fisher_validation_2007}. Many of the aforementioned decision-making scenarios involve high-stakes decisions, so it is vital to find alternative ways to communicate uncertainty estimates to people with low numeracy skills.

Besides numeracy skills, research shows that humans in general suffer from a variety of cognitive biases, some of which hinder our understanding of uncertainty~\citep{kahneman2011thinking,reyna_numeracy_2008, spiegelhalter_risk_2017}. One is called ratio bias, which refers to the phenomenon where people sometimes believe a ratio with a big numerator is larger than an equivalent ratio with a small numerator. For example, people may see 10/100 as a larger odds of having breast cancer than 1/10. This same phenomenon is sometimes manifested as an underweighting of the denominator, e.g. believing 9/11 is smaller than 10/13. This is also called denominator neglect.

In addition to ratio biases, people's perception of probabilities is also distorted in that they tend to underweight high probabilities while overweighting low probabilities. This distortion prevents people from making optimal decisions. \citet{zhang_ubiquitous_2012} showed that when people are asked to estimate probabilities or frequencies of events based on memory or visual observations, their estimates are distorted in a way that follows a log-odds transformation of the true probabilities. Research also found that this bias occurs when people are asked to make decisions under risk and that their decisions imply such distortions~\citep{tversky_advances_1992,zhang_designing_2015}. Therefore, when communicating probabilities, we need to be aware that people's perception of high risks may be lower than the actual risk, while that of low risks may be higher than actual.

A different kind of cognitive bias that impacts people's perception of uncertainty is framing~\citep{kahneman2011thinking}. Framing has to do with how information is contextualized. Typically, people prefer options with positive framing (e.g., a 80\% likelihood of surviving breast cancer) than an equivalent option with negative framing (e.g., a 20\% likelihood of dying from breast cancer). This bias has an effect on how people perceive uncertainty information. A remedy for this bias is to always describe the uncertainty of both positive and negative outcomes, rather than relying on the audience to infer what's left out of the description.

\subsection{Communication Methods}
Choosing the right communication methods can address some of the above issues. \citet{hullman2018pursuit} review methods for evaluating the success of uncertainty visualization.
\citet{van_der_bles_communicating_nodate} categorize the different ways of expressing uncertainty into nine groups with increasing precision, from explicitly denying that uncertainty exists to displaying a full probability distribution. While high-precision communication methods help experts understand the full scale of the uncertainty of the ML models, low precision methods can suffice for lay people, who may have potentially low numeracy skills. We now focus on the pros and cons of the four more precise methods of communicating uncertainty: 1) describing the degree of uncertainty using a predefined categorization, 2) describing a numerical range, 3) showing a summary of a distribution, and 4) showing a full probability distribution. The first two methods can be communicated verbally, while the last two often require visualizations.

\begin{figure*}[htb]
    \begin{subfigure}{0.2\textwidth}
        \includegraphics[width=\textwidth]{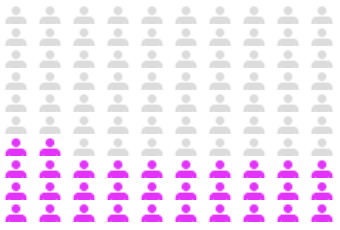}
        \caption{An example icon array chart. This can be used to represent the chance of a patient having breast cancer.}
        \label{fig:iconarray}
    \end{subfigure}
    \hfill
    \begin{subfigure}{0.25\textwidth}
         \centering
         \includegraphics[width=\textwidth]{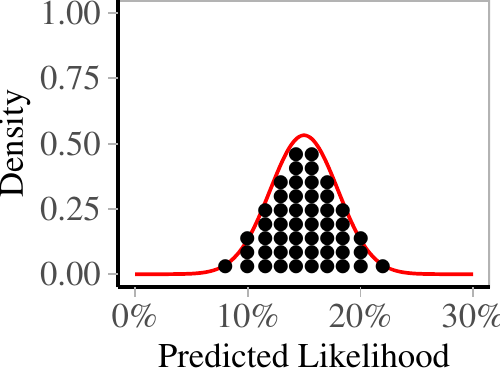}
         \caption{Quantile dot plot, which can be used to show the uncertainty around the predicted likelihood of a patient having breast cancer.}
         \label{fig:qdp}
     \end{subfigure}
     \hfill
     \begin{subfigure}{0.25\textwidth}
         \centering
         \includegraphics[width=0.85\textwidth]{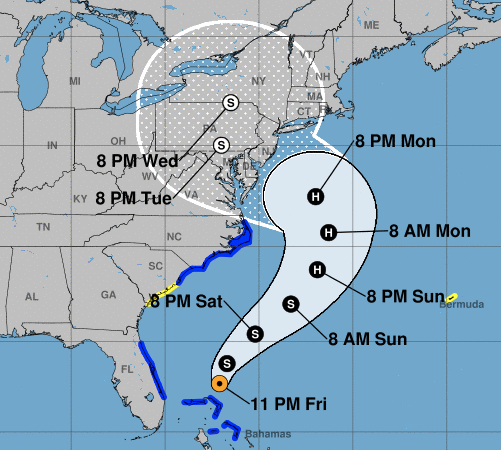}
         \caption{Cone of uncertainty, showing the uncertainty around the predicted path of the center of a hurricane. Taken from \cite{NHC_cone}.}
         \label{fig:cou}
     \end{subfigure}
     \hfill
     \begin{subfigure}{0.25\textwidth}
         \centering
         \includegraphics[width=\textwidth]{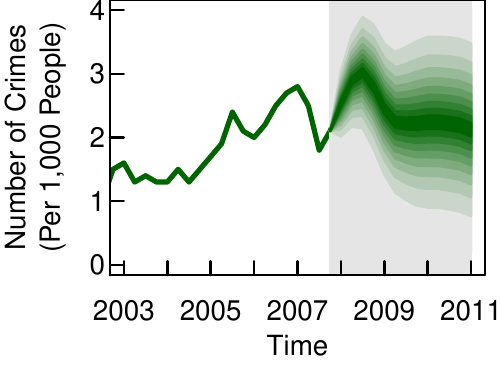}
         \caption{Fanchart, which can be used to show how the predicted crime rate of a city evolves over time.}
         \label{fig:fanchart}
     \end{subfigure}
        \caption{Examples of uncertainty visualizations.}
        \label{fig:three graphs}
\end{figure*}

A predefined, ordered categorization of uncertainty and risk levels reduces the cognitive effort needed to comprehend uncertainty estimates, and therefore is particularly likely to help people with low numeracy skills \citep{peters_numeracy_2007}. A great example of how to appropriately use this technique is the GRADE guidelines, which introduce a four-category system, from high to very low, to rate the quality of evidence for medical treatments~\citep{balshem_grade_2011}. GRADE has provided definitions for each category and a detailed description of the aspects of studies to evaluate for constructing quality ratings. Uncertainty ratings are also frequently used by financial agencies to communicate the overall risks associated with an investment instrument~\citep{dionisio2007entropy}.

The main drawback of communicating uncertainty via predefined categories is that the audience, especially non-experts, might not be aware of or even misinterpret the threshold criteria of the categories. Many studies have shown that although individuals have internally consistent interpretation of words for describing probabilities (e.g., likely, probably), these interpretations can vary substantially from one person to another~\citep{lichtenstein_empirical_1967,budescu_consistency_1985,clark_verbal_1990}. Recently, \citet{budescu_effective_2012} investigated how the general public interpret the uncertainty information in the climate change report published by the Intergovernmental Panel on Climate Change (IPCC). They found that people generally interpreted the IPCC's categorical description of probabilities as less likely than the IPCC intended. For example, people took the word ``very likely'' as indicating a probability of around 60\%, whereas the IPCC's guideline specifies that it indicates a greater than 90\% probability. To avoid such misinterpretation, categorical and numerical forms of uncertainty can be communicated when possible.

Though numbers and numerical ranges are more precise than categorical scales in communicating uncertainty, as discussed earlier, they are harder to understand for people with low numeracy and can induce ratio biases. However, a few techniques can be used to remediate these problems. First, to overcome the adverse effect of denominator neglect, it is important to present ratios with the same denominator so that they can be compared with just the numerator~\citep{spiegelhalter_visualizing_2011}. Denominators that are powers of 10 are preferred since they are easier to compute. There is so far no conclusive findings on whether frequency format (\textit{``Out of every 100 patients, 2 are likely to be misdiagnosed''}) is easier to understand than ratios/percentages (\textit{``Out of every 100 customers, 2\% are likely to be misdiagnosed''}): people do seem to perceive risk probabilities represented in the frequency format as showing higher risk than those represented in the percentage format~\citep{reyna_numeracy_2008}. Therefore, it is helpful to use a consistent format to represent probabilities. If the audience underestimates risk levels, the frequency format may be preferred.

Uncertainty estimates can also be represented with graphics, which have several advantages over verbal communication, such as attracting and holding the audience's attention, revealing trends or patterns in the data, and evoking mental mathematical operations \citep{lipkus_visual_1999}. Commonly used visualizations include pie charts, bar charts, and more recently, icon arrays (Figure~\ref{fig:iconarray}). Pie charts are particularly useful for conveying proportions since all possible outcomes are depicted explicitly. However, it is more difficult to make accurate comparisons with pie charts than with bar charts because pie charts use areas to represent probabilities. Icon arrays vividly depict part-to-whole relationship, and because they show the denominator explicitly, they can be used to overcome ratio biases.

So far, what we have discussed pertains mostly to conveying uncertainty of a binary event, which takes the form of a single number (probability), whereas the uncertainty of a continuous variable or model prediction takes the form of a distribution. This latter type of uncertainty estimate can be communicated either as a series of summary statistics about the distribution, or directly as the full distribution. As mentioned in Table~\ref{tab:uncertainty_metrics}, some commonly reported summary statistics include mean, median, confidence intervals, standard deviation, and quartiles \citep{theory_of_statistics}. These statistics are often depicted graphically as error bars and boxplots for univariate data, and two dimensional error bars and bagplots for bivariate data~\citep{rousseeuw_bagplot_1999}. We describe these summary statistics and plots in detail in the supplementary material.
Error bars only have a few graphical elements and are hence relatively easy to interpret. However, since they have represented a range of different statistics in the past, they are ambiguous if presented without explicit labeling~\citep{wilke_fundamentals_2019}. Error bars may also overly emphasize the range within the bar~\citep{correll_error_2014}. Boxplots and bagplots are less popular in the mass media, and generally require some training to understand.

When presenting uncertainty about a single model prediction, it might be better to show the entire posterior predictive distribution, which can avoid over-emphasis of the within-bar range and allow more granular visual inferences. Popular visualizations of distributions are histograms, density plots, and violin plots (\citet{hintze_violin_1998} show multiple density plots side-by-side), but they seem to be hard for an uninitiated audience to grasp. They are often mistaken as bivariate case-value plots in which the lines or bars denote values instead of frequencies \cite{boels_conceptual_2019}. More recently, \citet{kay_when_2016} develop quantile dot plots to convey distributions (see Figure~\ref{fig:qdp} for an example). These plots use stacked dots, where each dot represents a group of cases, to approximate the data frequency at particular values. This method translates the abstract concept of probability distribution into a set of discrete outcomes, which are more familiar concepts to people who have not been trained in statistics. \citet{kay_when_2016}'s study showed that people could more accurately derive probability estimates from quantile dot plots than from density plots. 
\citet{kruschke2014doing} defines a method that attempts to simultaneously convey aleatoric and epistemic uncertainty in a single plot by showing the predictive densities resulting from various samples the posterior distribution; however, the efficacy of such plots in practice is still unknown.

One very different approach to conveying uncertainty is to individually show random draws from the probability distribution as a series of animation frames called hypothetical outcome plots (HOP) \cite{hullman_hypothetical_2015}. Similar to the quantile dot plots, HOPs accommodate the \textit{frequency} view of uncertainty very well. In addition, showing events individually does not add any new visual encodings (such as the length of the bar or height of the stacked dots) and thus requires no additional learning from the viewers. \citet{hullman_hypothetical_2015} show that this visualization enabled people to make more accurate comparisons of two or three random variables than error bars and violin plots, presumably because statistical inference based on multiple distribution plots require special strategies while HOP does not. The drawbacks of HOP are: (a) it takes more time to show a representative sample of the distribution, and (b) it may incur high cognitive load since viewers need to mentally count and integrate frames. Nevertheless, because this method is easy to understand for people with low numeracy, similarly animated visualizations are frequently used in the mass media~\citep{yau_years_2015,badger_income_2018}. \citet{hofman2020visualizing} found that 95\% confidence intervals (inferential uncertainty) can be more misleading that 95\% prediction intervals (outcome uncertainty): HOPs tend to help reduce error when laypeople are asked to estimate effect size. However, \citet{kale2020visual} note that simple heuristics may suffice instead of complex uncertainty visualizations: they find that the best visualizations for understanding effect size may not be the best for decision-making.

The above methods are designed to communicate uncertainty around a single quantity, so they need to be extended for visualizing uncertainty around a range of predictions, such as those in time-series forecasting. The simplest form of such visualization is a quantile plot, which uses lines to connect predictions at equal quantiles of the uncertainty distribution across the output range. When used in time-series forecasting, such plots are called cone-of-uncertainty plots (see Figure~\ref{fig:cou}), in which the cone enlarges over time, indicating increasingly uncertain predictions. Gradient plots, or fan charts (see Figure~\ref{fig:fanchart}) in the context of time series forecasting, can be used to show more granular changes in uncertainty, but they require extra visual encoding that may not be easily understood by the viewer. In contrast, spaghetti plots simply represent each model's predictions as one line, while uncertainty  can be inferred from the closeness of the lines. However, they might put too much emphasis on the lines themselves and de-emphasize the range of the model predictions. Lastly, HOP can also be used to show uncertainty estimates over a range of predictions by showing each model's predictions in an animation frame~\citep{wilke_fundamentals_2019}.

\subsection{Uncertainty Requirements}\label{sec:requirements}
Familiarity with the findings above will be helpful for teams building uncertainty into ML workflows. Yet, none of these findings should be treated as conclusive when it comes to predicting how usable a given expression of uncertainty will be for different types of users facing different kinds of constraints in real-world settings. Instead, findings from the literature should be treated as fertile ground for generating hypotheses in need of testing with real users, ideally engaged in the concrete tasks of their typical workflow, and ideally doing so in real-world settings.

It is important to recognize just how diverse individual users are, and how different their social contexts can be. In our cancer diagnostic scenario, the needs and constraints of a doctor making a time-pressed and high-stakes medical decision using an ML-powered tool will likely be very different from those of a patient attempting to understand their diagnosis, and different again from those of an ML engineer reviewing model output in search of strategies for model improvement. Furthermore, if we zoom in on any one of these user populations, we still typically observe a tremendous diversity in skills, experience, environmental constraints, and so on. For example, among doctors, there can be big differences in terms of statistical literacy, openness to trusting ML-powered tools, time available to consume and decide on model output, and so on. These variations have important implications for designing effective tools.

To design and build an effective expression of uncertainty, we need to begin with an understanding of who the tool will be used by, what goal that user has, and what needs and constraints the user has. Frequently we also need to understand the organizational and social context in which a user is embedded. For example, to understand how an organization calculates and processes risk, which can influence the design of human-in-the-loop processes, automation, where thresholds are set, and so on. This point is not a new one, and it is by no means unique to the field of ML. User-centered design (UCD), human-computer interaction (HCI), user experience (UX), human factors, and related fields have arisen as responses to this challenge across a wide range of product and tool design contexts~\cite{goodman2009three,goodman2012observing,preece2015interaction}.

UCD and HCI have a firm footing in many software development contexts, yet they remain relatively neglected in the field of ML. Nevertheless, a growing body of research is beginning to demonstrate the importance of user-centered design for work on ML tools~ \citep{inkpen2019human,thieme2020machine}. For example, \citet{yang2016investigating} draw on field research with healthcare decision-makers to understand why an ML-powered tool that performed well in laboratory tests was rejected by clinicians in real-world settings. They found that users saw little need for the tool, lacked trust in its output, and faced environmental barriers that made it difficult to use. \citet{narayanan2018humans} conduct a series of user tests for explainability to uncover which kinds of increases in explanation complexity have the greatest effect on the time it takes for users to achieve certain tasks. \citet{doshi2017towards} propose a framework for evaluation of explainability that incorporates tests with users engaged in concrete and realistic tasks. From a practitioner's perspective, \citet{lovejoy2018ux} describes the user-centered design lessons learned by the team building Google Clips, an AI-enabled camera designed to capture candid photographs of familiar people and animals. One of their key conclusions is that ``[m]achine learning won't figure out what problems to solve. If you aren't aligned with a human need, you're just going to build a very powerful system to address a very small — or perhaps nonexistent — problem.''

Research to uncover user goals, needs, and constraints can involve a wide spectrum of methods, including but not limited to in-depth interviews, contextual inquiry, diary studies, card sorting studies, user tests, user journey mapping, and jobs-to-be-done workshops with users~\citep{rubin2008plan,goodman2009three,goodman2012observing}. It is helpful to divide user research into two buckets: 1) discovery research, which aims to understand what problem needs to be solved and for which type of user; and 2) evaluative research, which aims to understand how well our attempts to solve the given problem are succeeding with real users. Ideally, discovery research precedes any effort to build a solution, or at least occurs as early in the process as possible. Doing so helps the team focus on the right problem and right user type when considering possible solutions, and can help a team avoid costly investments that create little value for users. Which of the many methods a researcher uses in discovery and evaluative research will depend on many factors, including how easy it is to find relevant participants, how easy it is for the researcher to observe participants in the context of their day-to-day work, how expensive and time-consuming it is for teams to prototype potential solutions for the purposes of user testing, etc. One takeaway is that teams building uncertainty into ML workflows should do user research to understand what problem needs solving and for what type of user.

\bigskip
\noindent\textbf{Takeaways}: 
\begin{enumerate}
    \item Stakeholders struggle interpreting uncertainty estimates. 
    \item While uncertainty estimates can be represented with a variety of methods, the chosen method should be one that is tested with stakeholders.
    \item Teams integrating uncertainty into ML workflows should undertake user research to identify the problems they are solving for and cater to different stakeholder types. 
\end{enumerate}

\section{Conclusion}
Throughout this paper, we have argued that uncertainty is a form of transparency and is thus pertinent to the machine learning transparency community. We surveyed the machine learning, visualization/HCI, design,  decision-making, and fairness literature. We reviewed how to quantify uncertainty and leverage it in three use cases: (1) for developers reducing the unfairness of models, (2) for experts making decisions, and (3) for stakeholders placing their trust in ML models.
We then described the methods for and pitfalls of communicating uncertainty, concluding with a discussion on how to collect requirements for leveraging uncertainty in practice.
In summary, well-calibrated uncertainty estimates improve transparency for ML models. In addition to calibration, it is important that these estimates are applied coherently and communicated clearly to various stakeholders considering the use case at hand.
Future work could study the interplay between fairness, transparency, and uncertainty. For example, one could explore how communicating uncertainty to a stakeholder affects their perception of a model's fairness, or one could study how to best measure the calibration of uncertainty in regression settings.
We hope this work inspires others to study uncertainty as transparency and to be mindful of uncertainty's effects on models in deployment.

\section*{Acknowledgments}
The authors would like to thank the following individuals for their advice, contributions, and/or support: James Allingham (University of Cambridge), McKane Andrus (Partnership on AI), Kemi Bello (Partnership on AI), Hudson Hongo (Partnership on AI), Eric Horvitz (Microsoft), Jessica Hullman (Northwestern University), Matthew Kay (Northwestern University), Terah Lyons (Partnership on AI), Elena Spitzer (Google), Richard Tomsett (IBM), Kush Varshney (IBM), and Carroll Wainwright (Partnership on AI).

UB acknowledges support from DeepMind and the Leverhulme Trust via the Leverhulme Centre for the Future of Intelligence (CFI), and from the Partnership on AI. JA acknowledges support from Microsoft Research, through its PhD Scholarship Programme, and from the EPSRC. 
AW acknowledges support from a Turing AI Fellowship under grant EP/V025379/1, The Alan Turing Institute under EPSRC grant EP/N510129/1 and TU/B/000074, and the Leverhulme Trust via CFI.

\bibliographystyle{ACM-Reference-Format}
\bibliography{paper}

\clearpage
\appendix

\section{Uncertainty Quantification Metrics}
\label{uncmetrics}
In this appendix we detail different metrics with which the uncertainty in a predictive distribution can be summarized. We distinguish between metrics that communicate aleatoric uncertainty, those that communicate epistemic uncertainty, and those that inform us about the combination of both. We also distinguish between the classification and regression setting. Recall that, as discussed in Section~\ref{sec:communication}, predictive probabilities more intuitively communicate a model's predictions and uncertainty to stakeholders than a continuous predictive distribution. Therefore, summary statistics for uncertainty might play a larger role when building transparent regression systems.

\subsection{Classification setting}
\label{appdx:uncmetricsclassification}
\textbf{Notation:} Let $\mathcal{D}=\left\{\mathrm{x_i}, \mathrm{y_i}\right\}_{i=1}^{N}$ represent a dataset consisting of $\mathrm{N}$ samples, where for each example $\mathrm{i}$, $\mathrm{x_i} \in \mathcal{X}$ is the input and $\mathrm{y_{i}} \in \mathcal{Y}=\small{\{{c}_{k}}\}_{k=1}^{K}$ is the ground-truth class label. Let $p_{i}\left(\mathrm{y}| \mathrm{x}_{i}, \mathrm{w}\right)$ be the output from the parametric classifier $f_{\mathrm{\mathrm{w}}}\left(\mathrm{y} | \mathrm{x_i}\right)$ with model parameters $\mathrm{w}$. In probabilistic models, the output predictive distribution can be approximated from $\mathrm{T}$ stochastic forward passes (Monte Carlo samples), as described in Section~\ref{sec:measuring}: $p_i\left(\mathrm{y}| \mathrm{x_i}\right)=\frac{1}{T} \sum_{t=1}^T p_i^t\left(\mathrm{y} | \mathrm{x_i}, \mathrm{w}_{t}\right)$, where $\mathrm{w}_{t} \sim p(\mathrm{w}|\mathrm{D})$.

\paragraph{\textbf{Predictive entropy:}} The entropy~\cite{shannon1948mathematical} of the predictive distribution is given by Equation~\ref{eqn:pe}. Predictive entropy represents the overall predictive uncertainty of the model, a combination of aleatoric and epistemic uncertainties~\cite{mukhoti2018evaluating}. 

\begin{equation}
\small
\label{eqn:pe}
\mathbb{H}(\mathrm{y}|\mathrm{x},\mathrm{D}){:=}-\sum_{k=1}^{K}\left(\frac{1}{T} \sum_{t=1}^{T} a_t \right) \log \left(\frac{1}{T} \sum_{t=1}^{T} a_t \right)
\end{equation}
where $a_t = p\left(\mathrm{y}{=}{c}_{k} | \mathrm{x}, \mathrm{w}_{t}\right)$.In the case of point-estimate deterministic models, predictive entropy is given by Equation~\ref{eqn:pe_det} and captures only the aleatoric uncertainty.

\begin{equation}
\label{eqn:pe_det}
\mathbb{H}(\mathrm{y}|\mathrm{x},\mathrm{D}):=-\sum_{k=1}^{K} p\left(\mathrm{y}={c}_{k} | \mathrm{x}, \mathrm{w}\right) \log \left( p\left(\mathrm{y}={c}_{k} | \mathrm{x}, \mathrm{w}\right)\right)
\end{equation}

The predictive entropy always takes positive values between 0 and $\log K$. Its maximum is attained when the probability of all classes is $\frac{1}{K}$. Predictive entropy can be additively decomposed into aleatoric and epistemic components:

\paragraph{\textbf{Expected entropy:}} The expectation of entropy obtained from multiple stochastic forward passes captures the aleatoric uncertainty.

\begin{equation}
\small
\mathbb{E}_{p(\mathrm{w}|D)}\left[\mathbb{H}(\mathrm{y|x}, \mathrm{w})\right] {:=} \frac{1}{T} \sum_{t=1}^{T}\left( -\sum_{k=1}^{K} a_t \log(a_t) \right)
\label{eq:expectedentropy}
\end{equation}
\paragraph{\textbf{Mutual information:}} The mutual information~\cite{shannon1948mathematical} between the posterior of model parameters and the targets captures epistemic uncertainty~\cite{houlsby2011bayesian,gal2016uncertainty}. It is given by Equation~\ref{eq:mutualinformation}.

\begin{equation}
MI(\mathrm{y},\mathrm{w|x}, D) := \mathbb{H}(\mathrm{y}|\mathrm{x},\mathrm{D}) -\mathbb{E}_{p(\mathrm{w}|D)}\left[\mathbb{H}(\mathrm{y|x}, \mathrm{w})\right]\\
\label{eq:mutualinformation}
\end{equation}

The predictive entropy can be recovered as the addition of the expected entropy and mutual information: $$\mathbb{H}(\mathrm{y}|\mathrm{x},\mathrm{D}) = \mathbb{E}_{p(\mathrm{w}|D)}\left[\mathbb{H}(\mathrm{y|x}, \mathrm{w})\right] + MI(\mathrm{y},\mathrm{w|x}, D)$$

\paragraph{\textbf{Variation ratio:}} Variation ratio~\cite{freeman1965} captures the disagreement of a model's predictions across multiple stochastic forward passes given by Equation~\ref{eq:variationratio}.

\begin{equation}
VR := 1 - \frac{f}{T}
\label{eq:variationratio}
\end{equation}

where, f represents the number of times the output class was predicted from $\mathrm{T}$ stochastic forward passes. 

\subsection{Regression setting}
We assume the same notation as in the classification setting with the distinction that our targets $\mathrm{y_{i}} \in \mathcal{Y}\,{=}\,\mathcal{R}$ are continuous. For generality, we employ heteroscedastic noise models. Recall that this means our aleatoric uncertainty may be different in different regions of input space. We assume a Gaussian noise model with its mean and variance predicted by parametric models: $p(\mathrm{y} | \mathrm{x}, \mathrm{w}) = \mathcal{N}(\mathrm{y}; f^{\mu}_{\mathrm{w}}(\mathrm{x}), f^{\sigma^{2}}_{\mathrm{w}}(\mathrm{x}))$. Approximately marginalizing over $\mathrm{w}$ with T Monte Carlo samples induces a Gaussian mixture over outputs. Its mean is obtained as:
\begin{gather*}
    \mu \approx \frac{1}{T}\sum^{T}_{t=1} f^{\mu}_{\mathrm{w_{t}}}(\mathrm{x})
\end{gather*}
\paragraph{\textbf{Aleatoric and Epistemic Variances:}}
There is no closed-form expression for the entropy of the mixture of Gaussians (GMM). Instead, we use the variance of the GMM as an uncertainty metric. It also decomposes into aleatoric and epistemic components $(\sigma^{2}_{a}, \sigma^{2}_{e})$:
\begin{align*}
     \sigma^{2}(\mathrm{y} | \mathrm{x}, \mathrm{D})\,{=}\,\underbrace{\EX_{p(\mathrm{w} | \mathrm{D})}[f^{\sigma^{2}}_{\mathrm{w}}(\mathrm{y} | \mathrm{x})]}_{\sigma^{2}_{a}} + \underbrace{\sigma^{2}_{p(\mathrm{w} | \mathrm{D})}[f^{\mathrm{\mu}}_{\mathrm{w}}(\mathrm{y} | \mathrm{x})]}_{\sigma^{2}_{e}} .
\end{align*}
These are also estimated with MC:
\begin{align*}
    \sigma^{2}(\mathrm{y} | \mathrm{x}, \mathrm{D}) &\approx \underbrace{\frac{1}{T}\sum^{T}_{t=1} f^{\mu}_{\mathrm{w_{t}}}(\mathrm{y} | \mathrm{x})^{2} - (\frac{1}{T}\sum^{T}_{t=1} f^{\mu}_{\mathrm{w_{t}}}(\mathrm{y} | \mathrm{x}))^{2}}_{\sigma^{2}_{e}} + \\
    &+ \underbrace{\frac{1}{T}\sum^{T}_{t=1} f^{\sigma^{2}}_{\mathrm{w_{t}}}(\mathrm{y} | \mathrm{x})}_{\sigma^{2}_{a}}.
\end{align*}
Here, $\sigma^{2}_{e}$ reflects model uncertainty - our lack of knowledge about $\mathrm{w}$ - while $\sigma^{2}_{a}$ tells us about the irreducible uncertainty or noise in our training data.
Similarly to entropy, we can express uncertainty in regression as the addition of aleatoric and epistemic components. 

We now briefly discuss other common summary statistics to describe continuous predictive distributions \citep{theory_of_statistics}. Note that these do not admit simple aleatoric-epistemic decompositions.

\textbf{Percentiles:}
Percentiles tell us about the values below which there is a certain probability of our targets falling. For example, if the \nth{20} percentile of our predictive distribution is 5, this means that values $\leq 5$ take up $20\%$ of the probability mass of our predictive distribution.  

\textbf{Confidence Intervals:}
A confidence interval $[a, b]$ communicates that, with a probability $p$, our quantity of interest will lie within the provided range $[a, b]$ . Thus, the commonly used $95\%$ confidence interval tells us that percentile $2.5$ of our predictive distribution corresponds to $a$ and percentile $97.5$ corresponds to $b$.

\textbf{Quantiles:}
Quantiles divide the predictive distribution into sections of equal probability mass. 
Quartiles, which divide the predictive distribution into four parts, are the most common use of quantiles. The first quartile corresponds to the \nth{25} percentile, the second to the \nth{50} (or median) and the third to percentile 75.

\begin{figure}[]
    \vspace{-0.05in}
     \centering
     \includegraphics[width=\columnwidth]{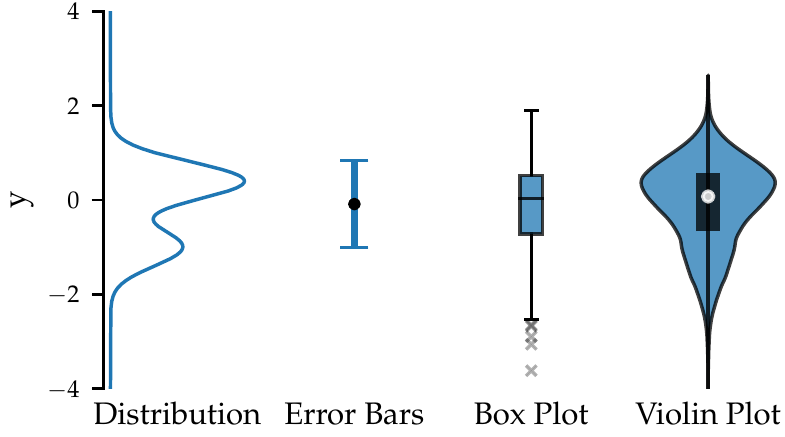}
     \vspace{-0.2in}
    \caption{Left: A full predictive distribution over some value of interest $y$. Center left: A summary of the predictive distribution composed of its mean value and standard deviation error-bars. Center right: A box plot showing quartiles and $1.5\times$ interquartile range whiskers. Right: A violin plot showing quartiles and sampled extrema.}
    \label{fig:regression_plots}
    \vspace{-0.1in}
\end{figure}  

The summary statistics above are often depicted in error-bar plots, box-plots, and violin plots, as shown in Figure~\ref{fig:regression_plots}. Error bar plots provide information about the spread of the predictive distribution. They can reflect variance (or standard deviation), percentiles, confidence intervals, quantiles, etc. Box plots tell us about our distribution's shape by depicting its quartiles. Additionally, box plots often depict longer error bars, referred to as ``whiskers,'' which tell us about the heaviness of our distribution's tails. Whiskers are most commonly chosen to be of length 1.5 $\times$ the interquartile range (Q1 - Q3) or extreme percentile values, e.g. 2-98. Samples that fall outside of the range depicted by whiskers are treated as outliers and plotted individually. Although less popular, violin plots have been gaining some traction for summarizing large groups of samples. Violin plots depict the estimated shape of the distribution of interest (usually by applying kernel density estimation to samples). They combine this with a box plot that provides information about quartiles. However, differently from regular box plots, violin plots' whiskers are often chosen to reflect the maximum and minimum values of the sampled population.

\section{Calibration Metrics}
\label{appdx:calmetrics}


This section describes existing metrics that reflect the calibration of predictive distributions for classification. There are no widely adopted calibration metrics for regression within the ML community. However, the use of calibration metrics for regression is common in other fields, such as econometrics. We discuss how these can be adapted to provide analogous information to popular ML classification calibration metrics. 
Calibration metrics should be computed on a validation set sampled independently from the data used to train the model being evaluated. In our cancer diagnosis scenario, this could mean collecting validation data from different hospitals than those used to collect the training data.


\textbf{Test Log Likelihood} (higher is better): This metric tells us about how probable it is that the validation targets were generated using the validation inputs and our model. It is a proper scoring rule \citep{gneiting2007strictly} that depends on both the accuracy of predictions and their uncertainty. We can employ it in both classification and regression settings. Log-likelihood is also the most commonly used training objective for neural networks. The popular classification cross-entropy and regression mean squared error objectives represent maximum log-likelihood objectives under categorical and unit variance Gaussian noise models, respectively \citep{bishop_pattern}.  

\textbf{Brier Score}~\cite{brier1950verification} (lower is better): Proper scoring rule that measures the accuracy of predictive probabilities in classification tasks. It is computed as the mean squared distance between predicted class probabilities and one-hot class labels:
\begin{gather*}
        \mathrm{BS} = \frac{1}{N} \sum^{N}_{n=1} \frac{1}{K}\sum^{K}_{k=1} (p(\mathrm{y}=c_{k} | \mathrm{x}, \mathrm{w}) - \ind [\mathrm{y}=c_{k}])^{2}
\end{gather*}
Unlike log-likelihood, Brier score is bounded from above. Erroneous predictions made with high confidence are penalised less by Brier score than by log-likelihood. This can avoid outliers or misclassified inputs from having a dominant effect on experimental results. On the other hand, it makes Brier score less sensitive.

\textbf{Expected Calibration Error (ECE)} ~\cite{naeini2015obtaining} (lower is better): This metric is popularly used to evaluate the calibration of deep classification neural networks. ECE measures the difference between predictive confidence and empirical accuracy in classification.
ECE segregates a model's predictions into bins depending on their predictive probability. In our example, this would mean grouping all patients who have been assigned a probability of having a cancer $\in [0, k)$ into a first bin, those who have been assigned probability $\in [k, 2k)$ into the second bin, etc. For each bin, calibration error is the difference between the proportion of patients who actually have the disease and the average of the probabilities assigned to that bin. This is illustrated in Figure \ref{fig:ensemble_calibration_plot}, where we use 10 bins. In this example, our model presents overconfidence in bins with $p<0.5$ and underconfidence otherwise.

After dividing the [0,1] range into a set of bins $\{B_{s}\}_{s=1}^{S}$, ECE weights the miscalibration in each bin by the number of points that fall into it $\abs{B_{s}}$:
    \begin{gather*}
        \mathrm{ECE} = \sum^{S}_{s=1} \frac{\abs{B_{s}}}{N} \abs{\mathrm{acc}(B_{s}) - \mathrm{conf}(B_{s}))}
    \end{gather*}
    Here, 
    \begin{gather*}
        \mathrm{acc}(B_{s})\,{=}\,\frac{1}{\abs{B_{s}}} \sum_{\mathrm{x} \in B_{s}} \ind[\mathrm{y}\,{=}\,\argmax_{c_{k}}p(\mathrm{y}| \mathrm{x}, \mathrm{w})]\quad \textrm{and}\\
        \mathrm{conf}(B_{s})\,{=}\, \frac{1}{\abs{B_{s}}} \sum_{\mathrm{x} \in B_{s}} \max_{c_{k}} p(\mathrm{y}| \mathrm{x}, \mathrm{w}).
    \end{gather*}
   
ECE is not a proper scoring rule. A perfect ECE score can be obtained by predicting the marginal distribution of class labels $p(\mathrm{y})$ for every input. A well-calibrated predictor with poor accuracy would obtain low log likelihood values (undesirable result) but also low ECE (desirable result).
Although ECE works well for binary classification, the naive adaption to the multi-class setting results in a disproportionate amount of class predictions being assigned to low probability bins, biasing results. 
\cite{nixon2019measuring} and \cite{kull2019beyond} propose alternatives that mitigate this issue. 

\textbf{Expected Uncertainty Calibration Error (UCE)} ~\cite{laves2019well} (lower is better): This metric measures the difference in expectation between a model's error and its uncertainty. The key difference from ECE is this metric quantifies model miscalibration with respect to predictive uncertainty (using a single uncertainty summary statistic, Appendix~\ref{appdx:uncmetricsclassification}). This differs from ECE, which quantifies the model miscalibration with respect to confidence (probability of predicted class).
\begin{equation}
    \label{eq:uce}
    \mathrm{UCE}=\sum_{s=1}^{S} \frac{\left|B_{s}\right|}{N}\left|\operatorname{err}\left(B_{s}\right)-\operatorname{uncert}\left(B_{s}\right)\right|
\end{equation}
Here, 
\begin{gather*}
        \mathrm{err}(B_{s})\,{=}\,\frac{1}{\abs{B_{s}}} \sum_{\mathrm{x} \in B_{s}} \ind[\mathrm{y}\,{\neq}\,\argmax_{c_{k}}p(\mathrm{y}| \mathrm{x}, \mathrm{w})]\quad \textrm{and}\\
        \operatorname{uncert}\left(B_{s}\right)=\frac{1}{\left|B_{s}\right|} \sum_{\mathrm{x} \in B_{s}} \tilde{\mathrm{u}}
\end{gather*}
where $\tilde{\mathrm{u}}\in[0,1]$ is a normalized uncertainty summary statistic (defined in Appendix~\ref{appdx:uncmetricsclassification}). 

\textbf{Conditional Probabilities for Uncertainty Evaluation}  ~\cite{mukhoti2018evaluating} (higher is better): Conditional probabilities {\textit{p(\text{accurate }|\text{ certain})}} and {\textit{p(\text{uncertain }|\text { inaccurate})}} have been proposed in ~\cite{mukhoti2018evaluating} to evaluate the quality of uncertainty estimates obtained from different probabilistic methods on semantic segmentation tasks, but can be used for any classification task.

{\textit{p(\text{accurate }|\text{ certain})}} measures the probability that the model is accurate on its output given that it is confident on the same. {\textit{p(\text{uncertain }|\text { inaccurate})}} measures the probability that the model is uncertain about its output given that it has made inaccurate prediction. Based on these two conditional probabilities, \textit{patch accuracy versus patch uncertainty} (PA$\mathrm{v}$PU) metric is defined as below.

\begin{gather*}
\label{eq:p_ac}
p(\mathrm{accurate | certain}) = \frac{\mathrm{n}_{AC}}{\mathrm{n}_{AC}+\mathrm{n}_{IC}} \\
p(\mathrm{uncertain | inaccurate}) = \frac{\mathrm{n}_{IU}}{\mathrm{n}_{IC}+\mathrm{n}_{IU}} \\
PAvPU = \frac{\mathrm{n}_{AC} + \mathrm{n}_{IU}}{\mathrm{n}_{AC}+\mathrm{n}_{IC}+\mathrm{n}_{AU}+\mathrm{n}_{IU}}
\end{gather*}

Here, $\mathrm{n}_{AC}$, $\mathrm{n}_{AU}$, $\mathrm{n}_{IC}$, $\mathrm{n}_{IU}$ are the number of predictions that are accurate and certain (AC), accurate and uncertain (AU), inaccurate and certain (IC), inacurate and uncertain (IU) respectively.

\textbf{Regression Calibration Metrics}: We can extend ECE to regression settings, while avoiding the pathologies described by \cite{nixon2019measuring}. We seek to assess how well our model's predictive distribution describes the residuals obtained on the test set. It is not straightforward to define bins, like in standard ECE, because our predictive distribution might not have finite support. We apply the cumulative density function (CDF) of our predictive distribution to our test targets. If the predictive distribution describes the targets well, the transformed distribution should resemble a uniform with support $[0, 1]$. This procedure is common for backtesting market risk models \citep{market_risk_book}.

\textbf{Regression  Calibration Error (RCE)} (lower is better):
To assess the global similarity between our targets' distribution and our predictive distribution, we separate the $[0, 1]$ interval into $S$ equal-sized bins $\{B_{s}\}_{s=1}^{S}$. We compute calibration error in each bin as the difference between the proportion of points that have fallen within that bin and $\frac{1}{S}$:
\begin{align*}
    \mathrm{RCE} = \sum^{S}_{s=1} \frac{\abs{B_{s}}}{N}\cdot \abs{\frac{1}{S} - \frac{\abs{B_{s}}}{N}} \\ \abs{B_{s}} = \sum_{n=1}^{N} \ind[ CDF_{p(y | \mathrm{x}^{(n)})}(y^{(n)}) \in B_{s}]
\end{align*}

\textbf{Tail Calibration Error (TCE)} (lower is better):
In cases of model misspecification, e.g. our noise model is Gaussian but our residuals are multimodal, RCE might become large due to this mismatch, even though the moments of our predictive distribution might be generally correct.
We can exclusively assess how well our model predicts extreme values with a ``frequency of tail losses'' approach \citep{tail_frequency_paper}. Only considering calibration at the tails of the predictive distribution allows us to ignore shape mismatch between the predictive distribution and the true distribution over targets. Instead, we focus on our model's capacity to predict on which inputs it is likely to make large mistakes. We specify two bins $\{B_{0}, B_{1}\}$, one at each tail end of our predictive distribution, and compute:
\begin{gather*}
    \mathrm{TCE} = \sum_{s=0}^{1} \frac{\abs{B_{s}}}{\abs{B_{0}} + \abs{B_{1}}} \cdot \abs{\frac{1}{\tau} - \frac{\abs{B_{s}}}{N}}; \\ \abs{B_{0}} = \sum_{n=1}^{N} \ind[ CDF_{p(y | \mathrm{x}^{(n)})}(y^{(n)}) < \tau]; \\
    \abs{B_{1}} = \sum_{n=1}^{N} \ind[ CDF_{p(y | \mathrm{x}^{(n)})}(y^{(n)}) \geq (1-\tau)]
\end{gather*}
We specify the tail range of our distribution by selecting $\tau$. Note that this is slightly different from \citet{tail_frequency_paper}, who uses a binomial test to assess whether a model's predictive distribution agrees with the distribution over targets in the tails.
RCE and TCE are not proper scoring rules. Additionally, they are only applicable to one-dimensional continuous target variables.

\section{Uncertainty Quantification Methods}
\label{appdx:bayesianmethods}

In this appendix, we describe a range of approaches to uncertainty quantification which were omitted from  Section~\ref{sec:measuring} for brevity. We focus mainly on approaches for obtaining uncertainty estimates from deep learning models but emphasize that the presented techniques often may be applied more generally. 


\subsection{Bayesian Methods}

\begin{figure}[]
    \vspace{-0.05in}
     \centering
     \includegraphics[width=\columnwidth]{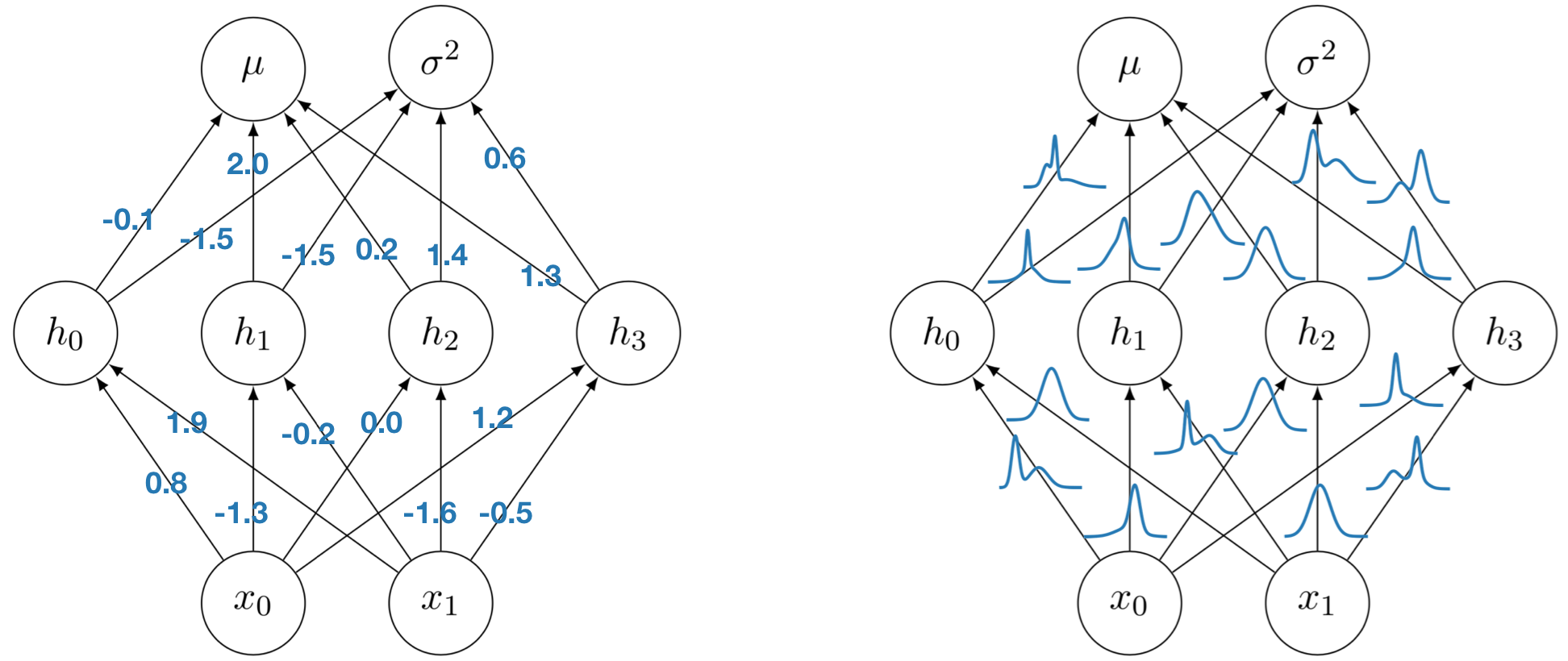}
     \vspace{-0.2in}
    \caption{Left: In a regular NN, weights take single values which are found by fitting the data with some optimisation procedure. In a Bayesian Neural Network (BNN), probability distributions are placed over weights. These are obtained by leveraging (or approximating) the Bayesian update~Eq(\ref{eq:bayes})\label{fig:NN_BNN}.}
    \vspace{-0.1in}
\end{figure}

Various approximate inference methods have been proposed for Bayesian uncertainty quantification in parametric models, such as deep neural networks. We refer to the resulting models as Bayesian Neural Networks (BNN), Figure~\ref{fig:NN_BNN}. Variational inference \cite{hinton1993keeping,graves2011practical,blundell2015weight,farquhar_radial_2020} approximates a complex probability distribution $p(\mathrm{w}|\mathrm{D})$ with a simpler distribution $q_\theta(\mathrm{w})$, parameterized by variational parameters $\theta$, by minimizing the Kullback-Leibler (KL) divergence between the two $KL(q_\theta(\mathrm{w}\,||\,p(\mathrm{w}|\mathrm{D}))$. In practice, this is done by maximising the evidence lower bound~(ELBO), as given by Equation~\ref{eqn:negELBO}.
 \begin{equation}
 \small
 \label{eqn:negELBO}
 \begin{aligned}
 \mathcal{L}_{\textrm{ELBO}} := -\mathbb{E}_{q_\theta(\mathrm{w})}\left[\log\,p(\mathrm{y}|\mathrm{x},\mathrm{w})\right] + KL[q_\theta(\mathrm{w})||p(\mathrm{w})]
 \end{aligned}
 \end{equation}
In Bayesian neural networks, this objective can be optimised with stochastic gradient descent optimization \citep{kingma2015variational}. After optimization $q_\theta(\mathrm{w})$ represents a distribution over plausible models which explain the data well. In mean-field variational inference, the approximate weight posterior $q_\theta(\mathrm{w})$ is represented by a fully factorized distribution, most often chosen to be Gaussian.  Some stochastic regularization techniques, originally designed to prevent overfitting, can also be interpreted as instances of variational inference. The popular Monte Carlo dropout~\cite{gal2016dropout} method is a form of variational inference which approximates the Bayesian posterior with a multiplicative Bernoulli distribution over sets of weights. The stochasticity introduced through minibatch sampling in batch-norm can also be seen in this light \citep{teye2018bayesian}. Stochastic weight averaging Gaussian (SWAG)~\cite{maddox2019simple} computes a Gaussian approximation to the posterior from checkpoints of a deterministic neural network's stochastic gradient descent trajectory.
These approaches are simple to implement but represent crude approximations to the Bayesian posterior. As such, the uncertainty estimates obtained by using these methods may suffer from some limitations \citep{foong_expressivity}.

Often, the predictive distribution, Equation~(\ref{eq:predicitve_posterior}), is also intractable.
In practice, for parametric models like NNs, it is approximated by making predictions with multiple plausible models, sampled from $p(\mathrm{w}|\mathrm{D})$. We see this in the leftmost plot from Figure~\ref{fig:ensemble_calibration_plot}: the predictions from different ensemble elements can be interpreted as samples which, when combined, approximate the predictive posterior \citep{wilson2020bayesian}. Different predictions tend to agree in the data-dense regions but they disagree elsewhere, yielding epistemic uncertainty.
Note, because we need to evaluate multiple models, sampling-based approximations of the predictive posterior incur additional computational cost at test time compared to non-probabilistic methods.

An alternative to variational inference is Stochastic gradient MCMC~\cite{welling2011bayesian,chen2014stochastic,zhang2020csgmcmc}. This class of methods allow us to draw biased samples from the posterior distribution over NN parameters in a mini-batch friendly manner.

Recent work has started to explore performing Bayesian inference over the function space of NNs directly.
\citet{sun2018functional} use a stochastic NN as a variational approximation to the posterior over functions. They define and approximately optimize a functional ELBO. \citet{Variational_Implicit_Processes} use a variant of the wake sleep algorithm to approximate the predictive posterior of a neural network with a Gaussian Process as a surrogate model. 
The Spectral-normalized Neural Gaussian Process (SNGP)~\cite{liu2020simple} enables us to compute predictive uncertainty through input distance awareness, avoiding Monte Carlo sampling.
Neural stochastic differential equation models (SDE-Net)~\cite{lingkai2020sdenet} provide ways to quantify uncertainties from the stochastic dynamical system perspective using Brownian motion. 
More recently, \citet{antoran2020depth} introduce Depth Uncertainty, an approach that captures model specification uncertainty instead of model uncertainty. By marginalizing over network depth their method is able to generate uncertainty estimates from a single forward pass.

Bayesian non-parametrics, such as Gaussian Processes \cite{gp_book}, are easy to deploy and allow for exact probabilistic reasoning, making their predictions and uncertainty estimates very robust. Unfortunately their computational cost grows cubically with the number of datapoints, making them a very strong choice of model only in the small data regime ($\leq$ 5000 points). Otherwise, approximate algorithms, such as variational inference \citep{matthews2017scalable}, are required. 

\subsection{Frequentist Methods}\label{appdx:non-bayesianmethods}

Within frequentist methods, post-hoc approaches are especially attractive; they allow us to obtain uncertainty estimates from non-probabilistic models, independently of how these were trained.
A principled way to do this is to leverage the curvature of a loss function around its optima. Optima in flatter regions suffer from less variance \cite{hochreiter1997flat}, suggesting our model's parameters are well specified by the data.
This is used to draw plausible weight samples by Resampling Uncertainty Estimation (RUE) \citep{schulam2019can} and local ensembles \citep{Madras2020Detecting}.
\citet{alaa2020discriminative} adapt the Jackknife, a traditional frequentist method, to generating post-hoc confidence intervals in small-scale neural networks. \cite{ahuja2019probabilistic} model the activations of neural networks at each hidden layer. They then check for distribution shift as mismatch between distributions learnt during training and test-time activations. 
    
Deterministic uncertainty quantification (DUQ)~\cite{van2020uncertainty} builds upon ideas of radial basis function networks~\cite{lecun1998gradient}, allowing one to obtain uncertainty, computed as the distance to a centroid in latent space, with a single forward pass. This same principle is leveraged by \cite{liu2020simple} to obtain epistemic uncertainty estimates.

\citet{guo2017calibration} show how using a validation set to learn a multiplicative scaling factor (known as a temperature) to output logits is a cheap post-hoc way to improve the calibration of NNs. 
Also worth mentioning is the orthonormal certificate method of \cite{tagasovska2019single}.

\section{Other Algorithmic Use Cases for Uncertainty}
Epistemic uncertainty can be used to guide model-based search in applications such as active learning \citep{houlsby2011bayesian,kirsch2019batchbald}.
In an active learning scenario, we are provided with a large amount of inputs but few or none of them are labelled. We assume labelling additional inputs has a large cost, e.g. querying a human medical professional. In order to train our model to make the best possible predictions, we would like to identify for which inputs it would be most useful to acquire labels. 
Here, each input's epistemic uncertainty tells us about how much our model would learn from seeing its labels and therefore directly answers our question.

Uncertainty can also be baked directly into the model learning procedure. In rejection-option classification, models can explicitly abstain from making predictions for points on which they expect to underperform~\cite{bartlett2008classification}. For example, our credit limit model may have high epistemic uncertainty when predicting on customers with specific attributes (e.g., extremely low incomes) that are underrepresented in the training data; as such, uncertainty can be leveraged to decide whether the model should defer to a human based on the observed income level.  Section~\ref{sec:use} reviews various ways to use uncertainty to improve ML models.

Distributional shift, which occurs when the distribution of the test set is different from the training set distribution, is sometimes considered as a special case of epistemic uncertainty, although \citet{malinin2018predictive} explicitly consider it as a third type of uncertainty: distributional uncertainty. \citet{ovadia2019can} compare different uncertainty methods in the specific case of dataset shifts. These techniques all have in common that they use epistemic uncertainty estimates to improve the model by identifying regions of the input space where the model performs badly due to a lack of data.
Some work also shows how using uncertainty techniques can directly improve the performance of a model and can out-perform using the soft-max probabilities \cite{gal2016dropout, kendall2015bayesian, kendall2016modelling}. \citet{miller2018dropout} discuss how dropout sampling helped in an open-set object detection task where new unknown objects can appear in the frame. Some other papers propose their own uncertainty techniques, and presents how they out-perform different baselines, often focusing on the out-of-distribution detection task \cite{devries2018learning, lakshminarayanan2017simple, malinin2018predictive}. 

It is also important to note that sometimes a model's output could be used for downstream decision-making tasks by another model, whether an ML model, or an operational research model which will optimize a plan based on a predicted class or quantity. Uncertainty estimates are essential to transparently inform the downstream models of the validity of the input. For instance, stochastic optimization can take distributional input to reduce the cost of the solution in the presence of uncertainty. More generally, in systems where models, humans, and/or heuristics are chained, it is crucial to understand how the uncertainty of each step interacts with each other, and how it impacts the overall uncertainty of the system.


\section{Fairness}

\subsection{Definitions}
\label{appdx:fairness}
Algorithmic fairness is complementary to algorithmic transparency. The ML community has attempted to define various notions of fairness statistically: see~\citet{barocas-hardt-narayanan} for an overview of the fairness literature. 
While there is no single definition of fairness for all contexts of deployment, many define ML fairness as absence of any prejudice or favoritism toward an individual or a group based on their inherent or acquired characteristics~\cite{Mehrabi2019ASO}. Unfairness can be the result of biases in (a) data used to build the algorithms or (b) the algorithms chosen to be implemented. 
We discuss some standard definitions of fairness here from the machine learning literature~\citep{hardt2016equality,beutel2017data}. Note that \citet{kleinberg2018inherent} finds that typically it is not possible to satisfy many fairness notions simultaneously. Let us assume we have a classifier $f$, which outputs a predicted outcome $\hat Y = f(X) \in \{0,1\}$ for some input $X$. Let $Y$ be the actual outcome for $X$. Let $A$ be a binary sensitive attribute that is contained explicitly in or encoded implicitly in $X$. When we refer to groups, we mean the two sets that result from partitioning a dataset $\mathcal{D}$ based on $A$ (i.e., if Group 1 was $\{X \in \mathcal{D} | A = 0 \}$, Group 2 would be $\{X \in \mathcal{D} | A = 1 \}$). Let $A = 0$ (Group 1) be considered unprivileged. Below are three common fairness metrics.
\begin{enumerate}
    \item Demographic Parity (DP): A classifier $f$ is considered to be fair with regard to DP (also known as statistical parity) if the following quantity is close to 0: $$P(\hat Y = 1 |A = 0) - P(\hat Y = 1 |A = 1)$$ The predicted positive rates for both groups should be the same~\citep{dwork2012fairness}.
    \item Equal Opportunity (EQ): A classifier $f$ is considered to be fair with regard to EQ if the following quantity is close to 0:  $$P(\hat Y = 1 |A = 0, Y = 1) - P(\hat Y = 1 |A = 1, Y = 1)$$ 
    That is, the true positive rates for both groups should the same~\citep{hardt2016equality}.
    \item Equalized Odds (EO): A classifier $f$ is considered to be fair with regard to EO if the following quantity is close to 0:  
    $$ \sum_{y \in \{0,1\}} |P(\hat Y = 1 |A = 0, Y = y) - P(\hat Y = 1 |A = 1, Y = y)|
    $$ 
    That is, we want to equalize the true positive and false positive rates across groups~\citep{hardt2016equality}. EO is satisfied if $\hat{Y}$ and ${A}$ are independent conditional on $Y$.
\end{enumerate}

\subsection{Uncertainty and Fairness}
\label{contd}
We also discuss the intersection of fariness and uncertainty, as it pertains to model specification and to bias mitigation
{\bf Uncertainty in model specification}
This type of uncertainty arises when the hypothesis class chosen for modeling the data may not contain the true data-generating process and could result in unwanted bias in model predictions.
For example, we might prefer a simple and explainable model for cancer diagnostics that, mistakenly, does not account for the non-linearity present in the data. 
Similarly, for ethical reasons we might choose to exclude the patient's race as a predictor from the model, when this information could help improve its performance~\cite{vyas2020hidden}.
The hypothesis class or the family of functions used to fit the data is primarily determined using domain knowledge and preferences of the model designer. For these reasons, the resulting model should be seen only as an approximation of the true data-generating process~\cite{buja2019models}.
In addition, using different benchmarks for the measurement of an algorithm's performance can lead to different choices in the final model. 
As a result, the trained classifier may not achieve high performance, even with potentially unlimited and rich data.

Bias arising from model uncertainty can potentially be mitigated by considering enlarging the hypothesis class considered, such as by using deep neural networks, when the datasets are sufficiently large. In general, this kind of bias is hard to disentangle from data uncertainty and therefore it can rarely be detected or analysed.

\noindent {\bf Uncertainty and bias mitigation}

We now present some of the methods to mitigate data bias and the possible implications of using uncertainty. These methods are typically categorized as pre-, in-, or post-processing based on the stage at which the model learning is intervened upon.

{\it Pre-processing} techniques modify the distribution of the data the classifier will be trained on, either directly~\cite{calmon2017optimized} or in a low-dimensional representation space~\cite{zemel2013learning}. Implicitly, these techniques reduce the uncertainty emerging from the features, outcome, or sensitive variables. 
Uncertainty estimation at this stage involves representing and comparing the training data distribution and random population samples with respect to target classes. 
The uncertainty measurements represented as distribution shifts of targeted classes between the training data and the population samples give a measure of training data bias which can be corrected by data augmentation of under-represented classes or by collecting larger and richer datasets~\cite{chen2018my}. The equalized odds post-processing method of~\citet{hardt2016equality} is guaranteed to reduce the bias of the classifier under an assumption on the noise in the sensitive attributes, namely the independence of the classifier prediction and the observed attribute, conditional on both the outcome and the true sensitive attribute~\cite{awasthi2020equalized}.

\textit{In-processing} methods modify the learning objective by introducing constraints through which the resulting classifier can achieve the desired fairness properties~\cite{agarwal2018reductions, zhang2018mitigating, donini2018empirical}.
Comparisons of the distribution of the features can be used to detect uncertainty in the model during training.
When little or no information about the sensitive attribute is available, distributionally robust optimization can be used to enforce fairness constraints~\cite{hashimoto2018fairness, wang2020robust}.
Algorithmic fairness approaches that employ active learning to either acquire features \cite{noriega2019active} or samples \cite{anahideh2020fair} with accuracy and fairness objectives also fall under this category.

\textit{Post-processing} techniques essentially modify the model's predictions post-training to satisfy a chosen fairness criterion~\cite{hardt2016equality}.
Frameworks like \cite{kamiran2012decision} can use this uncertainty information to de-bias the output of such models. Here, the predictions that are associated with high uncertainty can be skewed to favor a sensitive class as a de-biasing post-processing measure. Uncertainty in predictions can also be used to abstain from making decisions or defer the decisions to experts, which can lead to overall improvement in accuracy and fairness of the predictions \cite{madras2018predict}. An optimization-based approach is proposed in \cite{wei2019optimized} to transform the scores with suitable trade-off between utility of the predictions and fairness, with the assumption that the scores are well-calibrated.

Often there exist trade-offs between the different notions of fairness~\cite{corbett2017algorithmic, chouldechova2017fair, kleinberg2016inherent} (see Appendix \ref{appdx:fairness} for the definitions of the commonly used fairness metrics). 
For example, it has been shown that calibration and equalized odds cannot be achieved simultaneously when the base rates of the sensitive groups are different \cite{pleiss2017fairness}. 
Interestingly, this impossibility can be overcome, as shown in \cite{canetti2019soft}, by deferring uncertain predictions to experts. 
It is important that the uncertainty measurements produced by the prediction models are meaningful and unbiased \cite{romano2019malice} for reliable functioning of such bias mitigation methods and for communication to decision-makers.

\section{COVID Case Study}
\label{sec:covid}
\begin{figure}[]
    \vspace{-0.05in}
    \centering
    \includegraphics[width=0.95\linewidth]{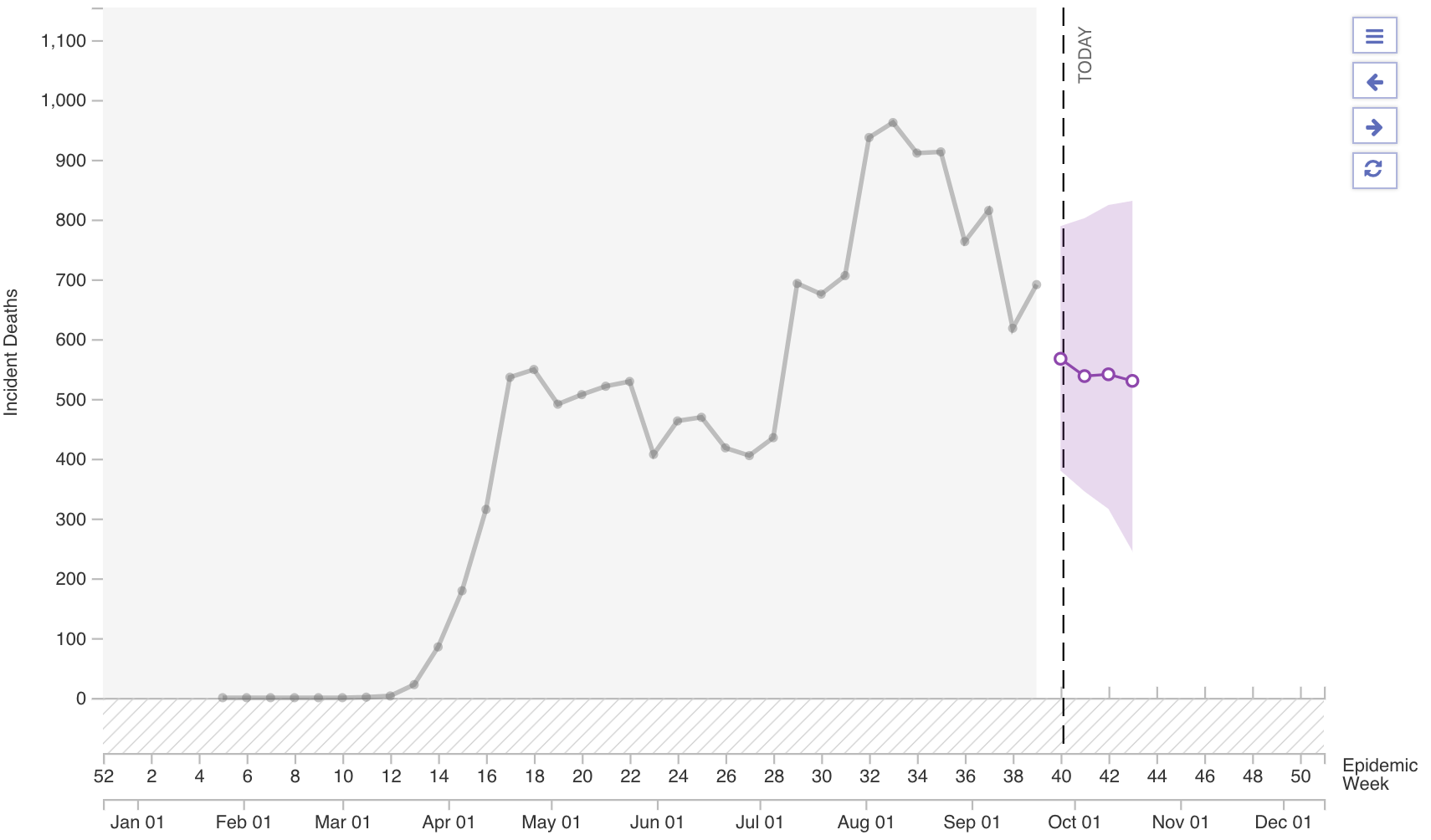}
    \caption{An example uncertainty visualization for projected COVID-19 mortality.}
    \label{fig:covid}
    \vspace{-0.15in}
\end{figure}
Uncertainty communication as a form of transparency is pivotal to garnering public trust during a pandemic. The COVID-19 global pandemic is an exemplar setting: disease forecasts, amongst other tools, have become critical for health communication efforts~\citep{jewell2020predictive}. In this setting, forecasts are being disseminated to governments, organizations, and individuals for policy, resource allocation, and personal-risk judgments and behavior~\citep{li2020estimated,eker2020validity,petropoulos2020forecasting}. The United States Centers for Disease Control and Prevention (CDC) maintains Influenza Forecasting Centers of Excellence, which have recently turned to creating public-facing hubs for COVID-19 forecasts, with the purpose of integrating infectious disease forecasting into public health decision-making~\citep{ray2020ensemble}. For example, we may want to forecast the number of deaths due to COVID-19. 
The CDC repository contains several individual models for this task. The types of models include the classic susceptible-infected-recovered infectious disease model, statistical models fit to case data, regression models with various types of regularization, and others. Each model has its own assumptions and approaches to computing and illustrating underlying predictive uncertainty. Figure~\ref{fig:covid} shows how uncertainty from one model is visualized as a predictive band with a mean estimate highlighted.
Given that such models are disseminated widely in the public via the Internet and other channels, and their result can directly affect personal behavior and disease transmission, this setting exemplifies an opportunity for user-centered design in uncertainty expression. In particular, assessing the forms of uncertainty visualization can be useful (e.g., 95\% confidence intervals versus 50\% confidence intervals, versus showing multiple different models, etc.). Indeed, the forms of uncertainty in COVID-19 forecasts could be used to inform a study to systematically assess user-specific uncertainty needs  (e.g., a municipal public health department may require conservative estimates for adequate resource allocation, while an individual may be more interested in trends to plan their own activities).

\end{document}